\documentclass[12pt]{iopart}
\usepackage{graphicx}
\usepackage{iopams}
\usepackage[utf8]{inputenc}
\usepackage{psfrag}
\usepackage{feynmp}
\newcommand{\E}[1]{\epsilon_{\bf #1}}
\newcommand{\feyndagg}[1]{/ \hspace{-2,5mm} #1}
\renewcommand{\Re}{ {\rm Re \  } }
\begin{document}

\title[Quark mass dependence of thermal excitations in one-loop QCD]
{Quark mass dependence of thermal excitations in QCD
in one-loop approximation}

\author{D.~Seipt$^1$, M.~Bluhm$^1$ and B.~K{\"a}mpfer$^{1,2}$}

\address{$^1$ Forschungszentrum Dresden-Rossendorf, PF 510119, 01314 Dresden, Germany}
\address{$^2$ Institut f{\"u}r Theoretische Physik, TU Dresden, 01062 Dresden, Germany}
\date{\today}

\begin{abstract}
A comprehensive determination of the quark mass dependence in the dispersion
relations of thermal excitations of gluons and quarks in non-Abelian gauge theory (QCD)
is presented for the one-loop approximation in Feynman gauge.
Larger values of the coupling are admitted,
and the gauge dependence is discussed. In a Dyson-Schwinger type approach, the effect
of higher orders is estimated for asymptotic thermal masses.
\end{abstract}
%\keywords{thermo-field theory, 
%self-energies, dispersion relations, one-loop approximation, quasi-particles}
%Use showkeys class option if keyword display desired

%PACS, the Physics and Astronomy Classification Scheme.
%\submitto{\jpg}

\maketitle

\section{Introduction}
The broad research programme of ultra-relativistic heavy-ion
collisions aims at investigating a new state of deconfined strongly
interacting matter. Experimentally, the hints for such a state,
dubbed quark-gluon plasma, have been accumulated in investigations at
CERN-SPS \cite{CERN-announcement} and further consolidated at
BNL-RHIC \cite{RHIC-announcement}. While early ideas have been guided
by asymptotic freedom considering the quark-gluon plasma as an ensemble
of weakly interacting quarks and gluons, the paradigm has
changed now to a strongly coupled quark-gluon plasma \cite{Shuryak}, as
enforced by the success of hydrodynamical concepts which point to an extremely fast
thermalization~\cite{Shuryak} and an
exceedingly low viscosity \cite{hydro,viscosity}.

Parallel to the
intense experimental efforts, which will soon proceed to a new era at
CERN-LHC, much progress has been achieved on the theoretical side.
In particular, first-principle (lattice) QCD calculations, directly based on
numerical evaluations of the QCD partition function, are progressing and
deliver information on the equation of state and related thermodynamic quantities
such as susceptibilities etc.\ \cite{lattice-QCD}. While final
results on bulk properties of deconfined strongly interacting
matter at finite temperatures seem to be achievable in the near future,
the microscopic nature -- with respect to the excitations in the considered medium -- received less attention
hitherto.

In fact, a variety of phenomenological approaches basing on different
microscopic pictures account fairly well for the bulk information obtained in lattice QCD.
For instance, various effective quasi-particle models
\cite{Peshier,Peshier2,Bluhm-PLB,Bluhm-PRC,Bluhm-EPJC,Biro,Toneev} rely on qualitatively
formulated / postulated excitations in the quark-gluon medium in order to
achieve a parametrization of the equation of state
adjusted to lattice QCD results. 
Certain symmetries of QCD are the basis of PNJL models (cf.~\cite{PNJL})
which also need adjustments to lattice QCD data.
Other approaches assume 
that the QGP may be described by a classical non-relativistic plasma 
of colour charges with Coulomb interaction \cite{Shuryak-Zahed} for temperatures $T_c<T<3T_c$,
where $T_c$ is the pseudo-critical or deconfinement temperature. 
Analytical attempts to derive the equation of state from QCD by
perturbative means \cite{Blaizot} are in agreement with the
non-perturbative lattice QCD results for temperatures $T > 3T_c$.
These approaches also provide some legitimation for a quasi-particle picture that
goes beyond strict perturbation theory \cite{blaizot-ipp}.

At asymptotically large temperatures suitable techniques like
dimensional reduction \cite{Rummukainen} bring further insight into the
structure of the theory and are applicable, e.g., to the physics of the early
universe. (Here, at temperatures 1 - 100 GeV, the heavy quark sector 
becomes important \cite{Kajantie}.)
Strict perturbative expansions have been pushed to order
$g^6 \ln g$ (see \cite{Rummukainen,Vuorinen} and references therein)
and slightly beyond \cite{Kajantie}, where $g$ is the strong coupling.
The successful hard thermal loop/hard dense loop (HTL/HDL) resummation scheme \cite{Braaten-Pisarski}
employs HTL self-energies, which correspond to the high temperature/density limit of one-loop self-energies, thus,
setting all quark masses to zero. The systematics of one-loop and HTL self-energies,
as well as the corresponding dispersion relations
were studied in \cite{Schertler} for zero quark masses.
Further steps towards going beyond the one-loop approximations have been attempted in
\cite{Nakkagawa}, where a Dyson-Schwinger scheme is set up in ladder approximation and in the chiral limit
focussing on the fermion spectral function.

It is the temperature region $T_c - 5 T_c$ which is of utmost
relevance for heavy-ion collisions. Here, the information on the
excitation spectrum of deconfined matter is fairly scarce. In
\cite{Petreczky}, the poles of quark and gluon propagators in Coulomb
gauge have been analyzed with the result that they may be
parametrized by an energy ($\omega$) - momentum ($k$) relation
$\omega^2=k^2+m_{q,g}^2$
with $m_{q,g} / T = \xi_{q,g}$ and $\xi_{q,g}$ ranging from 1.2 till 3.9
depending on parton species and temperature. 
Karsch and Kitazawa \cite{Karsch-Kitazawa} analyzed, 
in quenched approximation and in Landau gauge, spectral properties of quarks at $1.5 T_c$
and $3 T_c$ and at zero momentum as a function of the bare quark mass $m$. An
important result of these investigations is the confirmation of a mass gap $m_q = \xi_q(T) T$ with a
weak temperature dependence in the function $\xi_q(T)$. Another important result in \cite{Karsch-Kitazawa}
concerns the quark mass dependence of pole positions and residues of the quark propagator,
which was found to be qualitatively different from the expected perturbative pattern.

For the latter one, systematic investigations are hardly found in the literature.
The thermal self-energies of gluons (e.g.~\cite{Kalashnikov})
and quarks (e.g.~\cite{Petitgirard}) have been calculated for massive quarks.
Fermion mass effects on the dispersion relations have been studied by several authors
\cite{Petitgirard,Baym-Blaizot-Svetitsky,Blaizot-Ollitrault,Pisarski} with the restriction
to the long-wavelength limit of the dispersion relations or
to small (soft) masses $m \leq gT$ or $m \leq g\mu$, where $\mu$ is the quark chemical potential.

A more detailed investigation of the general mass dependence in the dispersion relations is desired, e.g., to uncover effects of masses $m \sim {\cal O} (T)$, 
as employed in lattice QCD calculations \cite{Allton}.
A deeper understanding of the impact of heavy quark masses on the equation of state
of strongly interacting matter relevant for the early universe is also of interest.

The chiral extrapolation, and thus the quark mass dependence, is an
important issue not only in thermo-field theory.
Also in low energy effective field theories, 
such as chiral perturbation theory, the chiral extrapolation of quantities 
like the nucleon mass is of great interest \cite{Weise}. 
In hot QCD it is particularly challenging as lattice QCD evaluations have been
performed often for nonphysically heavy quarks due to technical limitations. 
Apart from the necessary continuum extrapolation, the extrapolation to physical quark masses is an
inevitable step towards obtaining useable results.

Given this motivation, we present here the systematics of the quark
mass dependence of quasi-particle dispersion relations in hot one-loop QCD.
We mention as further motivation that in a series of papers
\cite{Peshier,Bluhm-PLB,Bluhm-PRC} the successful description of lattice QCD results on bulk
properties of strongly interacting matter was demonstrated
within a phenomenological quasi-particle model employing approximate
one-loop dispersion relations for quarks and gluons. 
The model \cite{Peshier,Bluhm-PLB,Bluhm-PRC,Bluhm-EPJC} 
goes beyond perturbation theory by using an
effective coupling rather than the strong coupling. This motivates us
not to restrict ourselves to the weak coupling regime.

Our paper is organized as follows. In section \ref{sec.21}, we present
numerical results of the thermal part of gluon one-loop self-energies and
related quantities, in particular the dispersion relations which offer
a glimpse on the relevant excitations. 
Section \ref{sec.22} illuminates the same quantities but now for quarks.
Section \ref{sec.3} is devoted to the
gauge dependence of our results. For this purpose, we 
contrast our results with calculations in Coulomb gauge.
Of course, there are regions in parameter space where our results
coincide with the HTL approximation, which has been proved
to be gauge independent in~\cite{Kunstatter-Kobes-Rebhan}.
The summary can be found in section \ref{sec.4}.
Appendix A lists necessary relations for the chiral expansion of the
asymptotic thermal masses and Appendix B contains the decomposition
of the quark propagator with non-zero quark mass into different physical excitations
according to the spinor structure of the propagator.
In Appendix C, figures for quantifying the self-energies are collected.
In Appendix D, we present an
asymptotic Dyson-Schwinger approach to estimate higher order contributions to the mass dependence
of the asymptotic thermal masses in the case of Abelian gauge theory.

\section{One-loop gluon excitations \label{sec.21}}

In the following, we consider the quark masses as well as the coupling $g$
as external parameters.
Our results are presented for two-flavor QCD ($N_f=2$) with one independent quark mass parameter to limit
the dimensions of parameter space. However, a generalization to cover a hierarchy of different quark masses
is straightforward.
We focus on the thermal parts of the self-energies at zero chemical potential
(quark--anti-quark symmetric hot QCD medium),
having in mind, however, the remarks given in \cite{Baym-Blaizot-Svetitsky}.

\subsection{Gluon self-energy}

The gluon one-loop self-energy may be decomposed as
\begin{equation}
\Pi_{\mu\nu,ab}^g(K) = \Pi_{\mu\nu,ab}^{YM}(K) + \sum_q \Pi_{\mu\nu,ab}^q(K)
\end{equation}
with Lorentz indices $\mu, \nu$ and colour indices $a, b$
of the adjoint representation of the gauge group. The contributions of
gluon loop ($3g$), gluon tadpole ($4g$), and ghost loop ($ghost$)
are grouped in the Yang-Mills contribution ''$YM$''. The quark loops are labelled by ''$q$'' and the sum
runs over all quark flavors included.
They read as functions of the four-momentum $K$ of the considered gluon
\begin{eqnarray}
\Pi{^{YM}_{\mu \nu,ab}}(K) &=&
\frac12  T \sum_{\omega_n}\int \frac{d^3 p}{(2\pi)^3}
\mbox{tr} \big[ \Gamma_{\mu \alpha \beta}^{3g} {\cal D}_{\alpha \sigma} (P)
                \Gamma_{\tau \sigma \nu}^{3g} {\cal D}_{\tau \beta} (Q) \big]_{ab}\nonumber \\
&+&
\frac12  T \sum_{\omega_n}\int \frac{d^3 p}{(2\pi)^3}
\mbox{tr} \big[ \Gamma_{\mu \alpha \beta \nu}^{4g} {\cal D}_{\alpha \beta} (P)\big]_{ab} 
\nonumber \\
&+&  T \sum_{\omega_n}\int \frac{d^3 p}{(2\pi)^3}
\mbox{tr} \big[ \Gamma_\mu^{ghost} G(P) \Gamma_\nu^{ghost} G(Q) \big]_{ab}\,\, , \\
\Pi{^{q}_{\mu \nu,ab}}(K)
&=&
- T \sum_{\omega_n}\int \frac{d^3 p}{(2\pi)^3}
\mbox{tr} \big[ \Gamma_{\mu}^{q} {\cal S}(Q) \Gamma_{\nu}^{q} {\cal S}(P) \big]_{ab}.
\end{eqnarray}
In the above expressions the fermion (F) and boson (B) propagators are
the bare propagators ${\cal S} ={\cal S}_0(K) = (m-\feyndagg{K}) \Delta_F (K)$ and
${\cal D}_{\mu\nu} = ({\cal D}_0)_{\mu \nu} = \delta_{\mu \nu} \Delta_B(K)$
in Feynman gauge suppressing the explicit colour indices for notational convenience, with the thermal propagators
$\Delta_F(K) = (\omega_n^2 + k^2 + m^2)^{-1}$ and
$\Delta_B(K) = (\omega_n^2 + k^2)^{-1}$ with
Matsubara frequencies $\omega_n = 2n \pi T$ and $\omega_n = (2n+1) \pi T$
for bosonic and fermionic degrees of freedom, respectively.
The trace ''tr'' has to be taken over spinor and colour indices and $P$ and $Q = P - K$ are the internal
loop four-momenta.
For the expressions of the various vertices $\Gamma$ we refer
to standard textbooks on thermo-field theory, e.g.~\cite{LeBellac}.

The tensor $\Pi_{\mu\nu}^g$ consists of two independent scalar functions according to
\begin{equation}
\Pi_{\mu\nu}^g = {\mathfrak P}_{\mu\nu}^T \Pi_T^g + {\mathfrak P}_{\mu\nu}^L \Pi_L^g,
\label{eq:decomposition_Pi}
\end{equation}
where the utilized four-transverse projectors ($K^\mu {\mathfrak P}_{\mu\nu}^{T,L}=0$) are given by
\begin{equation}
{\mathfrak P}_{\mu\nu}^T = g_{\mu\nu} - \frac{K_\mu K_\nu}{K^{2}} + \frac{{\mathfrak P}_{\mu\nu}^L}{N^{2}}, \qquad
{\mathfrak P}_{\mu\nu}^L = - N_\mu N_\nu
\end{equation}
with
$N_\mu = \left(K_\mu (K u)-u_\mu K^2\right) \left((K u)^2-K^2 \right)^{-1}$
and $K^\mu N_\mu = 0$.
The superscripts $T$ and $L$ indicate that the tensors ${\mathfrak P}^T_{\mu\nu}$ and
${\mathfrak P}^L_{\mu\nu}$ project on subspaces
transverse and longitudinal to the three-momentum ${\bf k}$, respectively.
After performing the summation over the Matsubara frequencies, integrating over the angular part of $d^3p$ and analytically continuing into Minkowski space-time one obtains for the individual scalar functions~\cite{Kalashnikov}
\begin{eqnarray}
\fl \Pi_L^{YM}(\omega,k)
&=& -\frac{C_A g^2}{\pi^2} \int \limits_0^\infty d p p \,n_B(p)
\left[ 1 - \frac{2k^2 - \omega^2 - 4p^2}{8pk} \ln {\cal A}_g + 
\frac{\omega}{2k} \ln {\cal B}_g \right], \\
\fl \Pi_T^{YM}(\omega,k)
&=& \frac{C_A g^2}{2\pi^2} \int \limits_0^\infty d p p\, n_B(p)
\left[1 + \frac{\omega^2}{k^2} - \frac{3k^2 + \omega^2 + 4p^2}{8pk^3}
(k^2-\omega^2) \ln {\cal A}_g \right. \nonumber \\
\fl & &  \left. - \frac{\omega}{2k^3}(k^2-\omega^2)  \ln {\cal B}_{g}\ \right], \\
\fl \Pi_{L}^q (\omega,k) &=& -\frac{2 C_2 g^2}{\pi^2}
\int \limits_0^\infty d p \,\frac{p^2}{\E{p}} n_F(\E{p})
\left[ 1 - \frac{k^2 - \omega^2 - 4\E{p}^2}{8k p} \ln {\cal A}_f
- \frac{\omega\E{p}}{2k p} \ln {\cal B}_f \right], \\
\fl \Pi_T^q (\omega,k)
&=& \frac{C_2 g^2}{\pi^2} \int \limits_0^\infty d p \,\frac{p^2}{\E{p}} n_F(\E{p})
\left[ 1+\frac{\omega^2}{k^2} -
\frac{k^4 - \omega^4 - 4\E{p}^2\omega^2 + 4p^2k^2}{8p k^3}
\ln {\cal A}_f \right. \nonumber \\
\fl & & + \left. \frac{\omega\E{p}}{2p k^3}(k^2-\omega^2) \ln {\cal B}_f \right]
\label{eq:photon_oneloop_tr}
\end{eqnarray}
with
\begin{eqnarray}
\fl {\cal A}_g &=& \frac{(k^2-\omega^2+2pk)^2 - 4p^2\omega^2}{(k^2-\omega^2-2pk)^2 - 4p^2\omega^2},
\qquad
{\cal B}_g = \frac{(k^2-\omega^2)^2-4p^2(k+\omega)^2}{(k^2-\omega^2)^2-4p^2(k-\omega)^2}, \\
\fl {\cal A}_f & = & \frac{(k^2 - \omega^2 + 2pk)^2-4\omega^2\E{p}^2}
{(k^2 - \omega^2 - 2pk)^2-4\omega^2\E{p}^2},
\qquad
{\cal B}_f = \frac{(k^2-\omega^2)^2 - 
4(pk + \omega\E{p})^2}{(k^2-\omega^2)^2 - 4(pk - \omega\E{p})^2}
\end{eqnarray}
and $\E{p} = \sqrt{p^2 + m^2}$.
$K \sim (\omega, {\bf k})$ consists of the two components
$\omega = K \cdot u$ and $\bf k$ with $|{\bf k}| = k = \sqrt{(K\cdot u)^2 - K^2}$
in the thermal medium with four-velocity $u_\mu$.  $n_{F,B}$ are the standard Fermi and Bose distribution functions
and the gauge group factors are $C_A = N_c$, $C_2 = 1/2$ for a $\mbox{SU}(N_c)$ gauge group with $N_c = 3$ for QCD.

\begin{figure}[t]
\include{small_frag}
\psfrag{xlabel}{\hspace*{-5mm} \small $\omega^2/g^2T^2$}
\psfrag{ylabel}{\hspace*{-6mm} \small $\Re \Pi_T^q/g^2T^2$}
\psfrag{label1}{}
\psfrag{label2}{}
\psfrag{label3}{$\omega^2-k^2$}
\begin{center}
\includegraphics[scale=0.32,angle=-90]{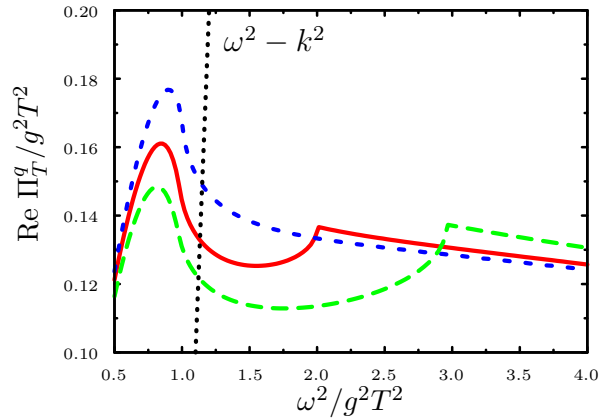} 
\end{center}
\caption{The scaled real part of the quark loop contribution to the transverse gluon 
self-energy according to (\ref{eq:photon_oneloop_tr}) as a function of the scaled energy
for $N_f=2$ degenerate quark flavors with mass $m/T = 0$ (blue short-dashed curve), $0.5$ (red solid curve) and
$0.7$ (green long-dashed curve). The kinky structure at $\omega = \sqrt{k^2+4m^2}$ 
is clearly visible. 
The black dotted line represents $\omega^2-k^2$. 
Thus, the intersection with the self-energy function would represent 
the sole solution of the dispersion relation
if there were no other contributions.
The curves are for $g = 1$ and $k = gT$.\label{fig.1}}
\end{figure}

For illustration, figure \ref{fig.1} exhibits the real part of $\Pi_T^q$ as a function of $\omega$ 
for three different values of the quark mass at fixed $k$.
The real parts of the quark loop contributions $\Pi_{T,L}^q$ develop a non-analytical 
(continuous but not continuously differentiable)
behaviour at the threshold  $\omega^2 = k^2 +4 m^2$ for the production 
of a real quark--anti-quark pair.
This behaviour is, however, related to the imaginary part of $\Pi_{T}^q$, which is also non-zero above
the threshold $\omega^2 > k^2 + 4 m^2$.
The imaginary parts in such one-loop calculations are known to have no direct physical meaning
since the quasi-particle damping rates (related to the widths of quasi-particle peaks in the spectral function)
turn out to be gauge dependent, both, in magnitude and even in sign.
Braaten and Pisarski~\cite{Braaten-Pisarski} pointed out the necessity of resumming
diagrams of higher-loop order (HTL resummation) to get a definite result.
However, the excitation energies of the quasi-particles considered in this paper
are rather independent of the imaginary parts.

With the replacements $g \to e$, $m \to m_e$, $C_A \to 0$, $C_2 \to 1$ and $N_f \to 1$
the quantities $\Pi_{T,L}^q$ would represent the thermal transverse and longitudinal one-loop photon self-energy
contributions in a hot Abelian (QED) plasma, respectively.
Such plasmas
are presently of interest \cite{Rafelski,Thoma}, as their
production under laboratory conditions seems feasible with high-intensity lasers.
The temperature of such a plasma is expected to be in the range of $T\sim 1\ldots10$ MeV
while the electron mass $m_e \approx 511$ keV
sets an additional energy scale. Thus, in this temperature region, in particular for lower temperatures,
the ratio $m_e/T$ is not small, which might lead to modifications in certain plasma properties as compared to the ultra-relativistic case. Consequently, a non-negligible electron mass is expected to influence physical observables
such as production rates of pions and muons \cite{Rafelski} and of photons \cite{Munshi}.

\subsection{Gluon dispersion relations}
\label{sect:2b}

The dressed gluon propagator may be decomposed as 
\begin{eqnarray}
{\cal D}_{\mu\nu}^{ab}(K) = \delta^{ab} \left(
{\mathfrak P}_{\mu\nu}^T \Delta_T + {\mathfrak P}_{\mu\nu}^L \frac{k^4}{K^4} \Delta_L +
\rho \frac{K_\mu K_\nu}{K^4} \right)
\end{eqnarray}
with gauge fixing parameter $\rho = 1$ in Feynman gauge
and the projectors ${\mathfrak P}_{\mu\nu}^{T,L}$ are defined in the explanation below equation (\ref{eq:decomposition_Pi}).
The transverse and longitudinal propagator parts are related to the corresponding self-energies
by Dyson's equation via
\begin{eqnarray}
\Delta_T = \frac{1}{K^2-\Pi^g_T(\omega, k)}, & \qquad &
\Delta_L = \frac{1}{k^2-\Pi^g_L(\omega, k)}
\end{eqnarray}
with $\Pi^g_{T,L} = \Pi_{T,L}^{YM} + \sum_q \Pi_{T,L}^q$.

\begin{figure}[ht]
% \psfrag{g = 0.3}{\tiny $g=0.3$}
% \psfrag{g = 1.0}{\tiny $g=1.0$}
% \psfrag{g = 3.0}{\tiny $g=3.0$}
% \psfrag{x}{\small $k/gT$}
% \psfrag{y}{\small $m/T$}
% \psfrag{G}{\hspace*{-1mm}\small ${\cal G}_T$}
% \psfrag{G2}{\hspace*{6cm} \small ${\cal G}_L$}
\begin{center}
 \includegraphics[scale=0.65]{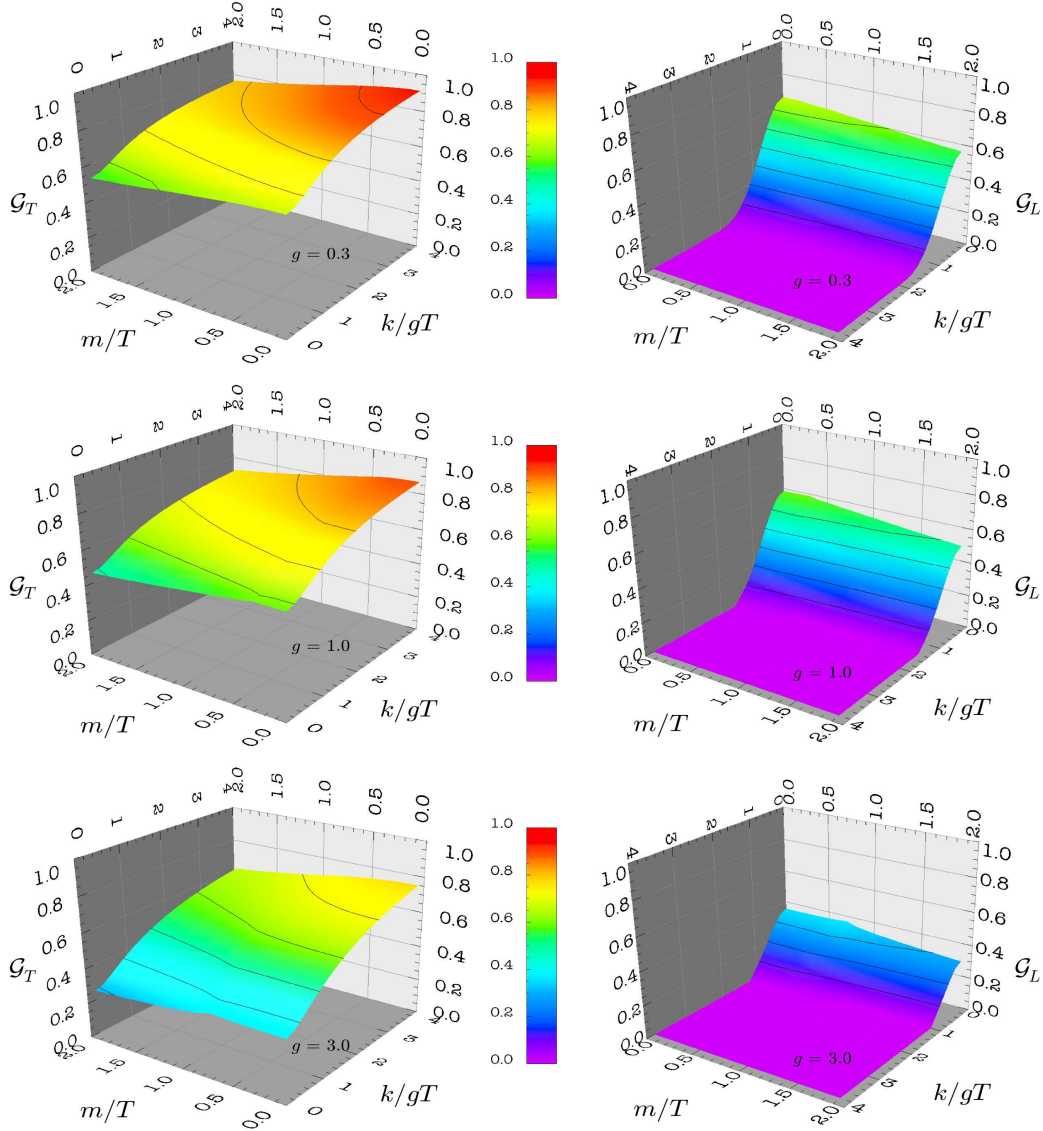} 
\end{center}
\caption{Transverse (left) and longitudinal (right) gluon dispersion relations 
represented as
{${{\cal G}_{T,L}} = ({\omega^2_{T,L}-k^2})/{m_g^2}$} as
functions of scaled momentum and scaled quark mass for 
$g = 0.3, \, 1.0, \, 3.0$ from top to bottom.
For a better presentation, the transverse and longitudinal quantities
are considered from different perspectives.
Faint isolines are given at ${\cal G}_{T,L} = 0.1,0.2,\ldots,0.9.$
\label{fig.2}}
\end{figure}

The gluon one-loop self-energies possess in general real and imaginary 
contributions. Here, we are interested in the dispersion relations
of gluonic quasi-particle excitations with negligible damping rates, which are characterized by
\begin{equation}
{\Re} \Big( \Delta_{T,L}^{-1} \Big) = 0. \label{eq:gluon_dispersion}
\end{equation}
The gluon excitation energies, which are the real solutions of (\ref{eq:gluon_dispersion}), may be represented by $\omega^2_{T,L} = k^2 +  {\cal G}_{T,L} m_g^2$ with $m_g^2 = \frac{1}{6}g^2T^2(N_c+N_f/2)$.
The contribution $m_g^2 {\cal G}_{T,L}$ to the quasi-particle excitation energy encodes the effects of the thermal medium and represents a thermal mass. Therefore, the light cone is given by the planes ${\cal G}_{T,L} \equiv 0$ in this representation.
For the transverse gluon
excitation one can write ${\cal G}_T \equiv \Re \Pi_T(\omega(k),k,m)/m_g^2$ and for longitudinal gluons (plasmons) one finds ${\cal G}_L \equiv K^2 \Re \Pi_L(\omega(k),k,m) / (k^2m_g^2)$.
In the presently considered case, the parameter space is three-dimensional for fixed temperature: The functions ${\cal G}_{T,L}$
depend on $m$, $g$ and $k$.
Note that the $g$ dependence cannot be scaled out, unlike in the HTL approximation. We are interested in the
regions $m \sim T$ and $k \sim T$. The first range is determined by the
lattice calculations \cite{Allton} which employ ''lattice quark masses''
$m \propto T$; the interesting range is extended till the chiral limit
$m = 0$. The second range is determined by the observation \cite{Peshier}
that thermal excitations with $k \sim T$ essentially contribute
to thermodynamic quantities.
A survey of these functions is exhibited in figure \ref{fig.2} for $N_f =2$ quarks of mass $m$.

The overall observation is, that the thermal contribution to the transverse gluon mode stays above the light cone in the whole parameter range, whereas it drops and rapidly approaches the light cone for the longitudinal mode at sufficiently large momenta, independent of $g$ and $m$.
Both, ${\cal G}_T$ and ${\cal G}_L$, are monotonically decreasing with increasing mass $m$, indicating that heavy quarks, compared to massless quarks, contribute not as much to the thermal mass of the gluon. The $m$ dependence is, though, rather weak because the thermal masses are dominated by the strong gluonic self-interaction.
For the transverse mode, the function ${\cal G}_T$ approaches unity in the limit $k \to \infty$ and $m \to 0$,
exposing that $m_g$ is the asymptotic mass, indeed.
We emphasize the strong $g$ dependence when considering the region
$g \sim 1$, see figure \ref{fig.2}. (Note that $g=1$ corresponds to $\alpha_S \approx 0.08$, whereas $g=0.3$ translates to $\alpha_S \approx {1}/{137}$ and $g=3$ means $\alpha_S \approx 0.72$). Clearly, for large values of $g$, higher order contributions are expected to become important.

Such 3D
plots are useful for surveys, however, cuts allow for a better quantitative
representation of results. Corresponding figures are relegated to Appendix C
(cf.~figures \ref{fig.5} and \ref{fig.6}).

\subsection{Gluon plasma frequency and asymptotic behaviour}

\begin{figure}[t]
\include{small_frag}
\psfrag{xlabel}{\small $g$}
\psfrag{ylabel}{\hspace*{-2.5mm}\small $\chi(g,m)$}
\psfrag{xlabelmass}{\hspace*{2.5mm}\small $m/T$}
\psfrag{ylabelmass}{\hspace*{2.5mm}\small $\cal I,J$}
\psfrag{labelI}{\tiny $\cal I$}
\psfrag{labelJ}{\tiny $\cal J$}
\includegraphics[scale=0.32,angle=-90]{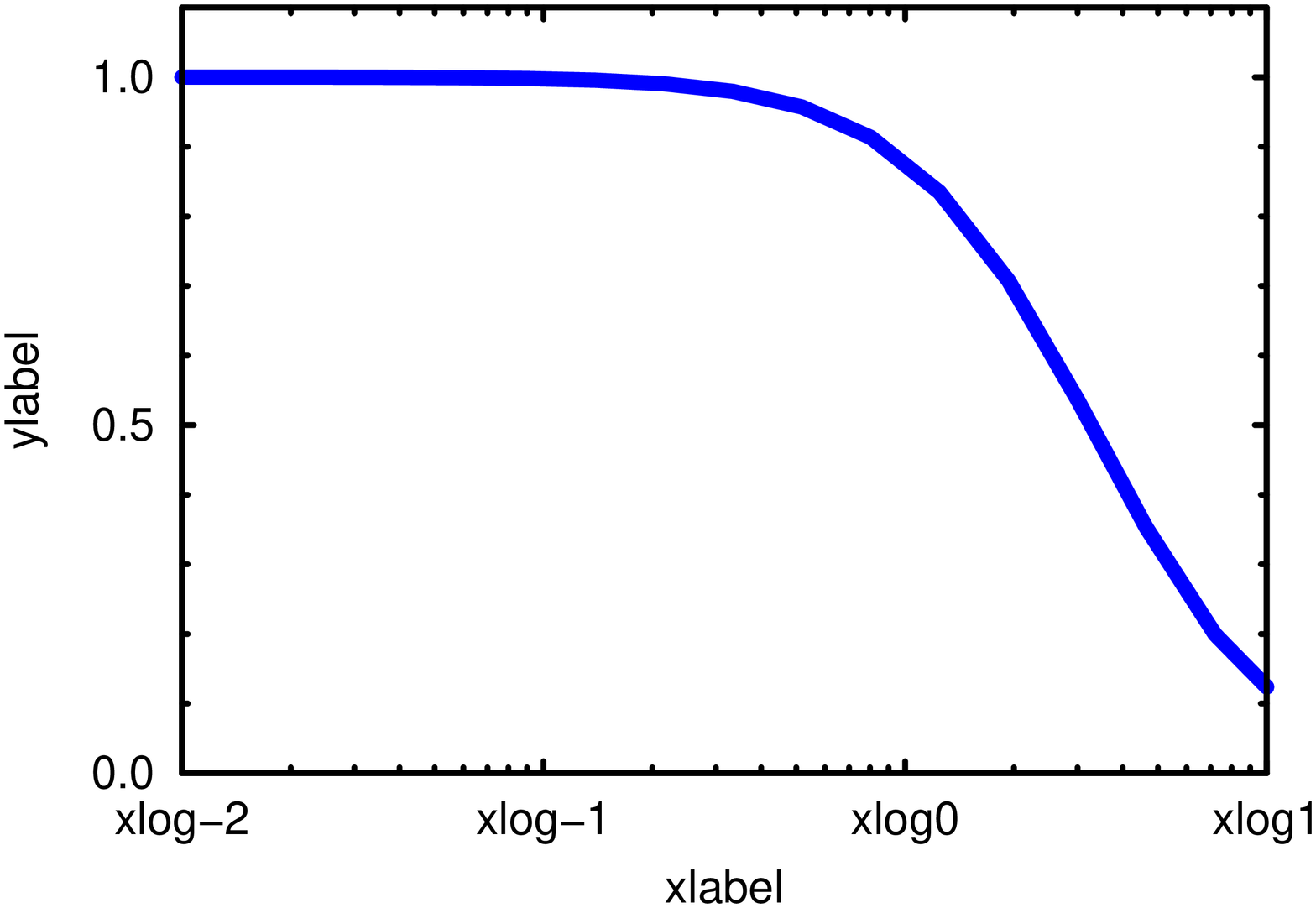}
\includegraphics[scale=0.32,angle=-90]{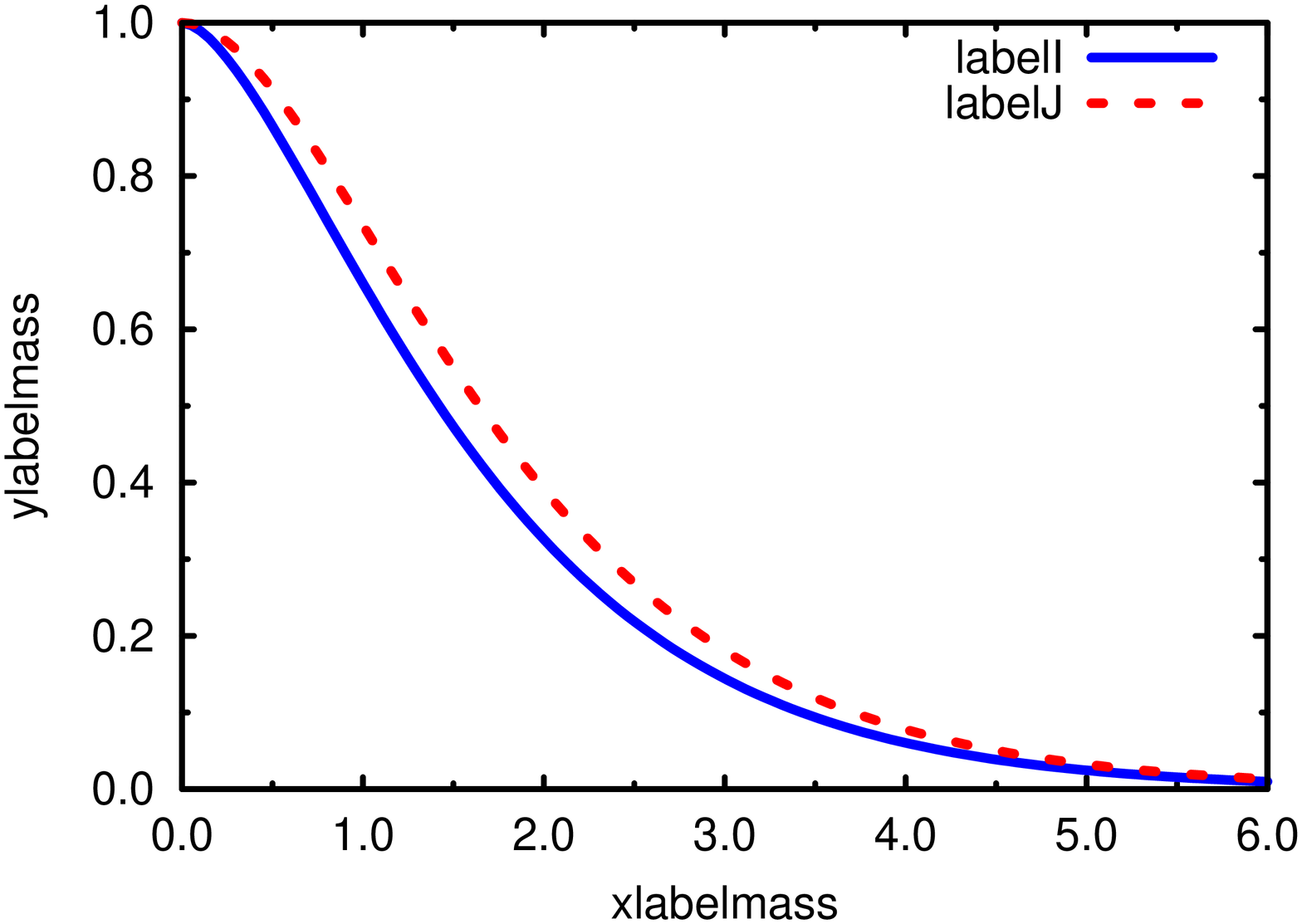}
\caption{Left panel: Function $\chi(g,m)$ defined in (\ref{eq:gluon_plasma_frequency}), 
showing the non-trivial dependence of the gluon plasma frequency 
as a function of the coupling $g$. On the displayed scale,
curves for $0 \leq m/T \leq 2$ are on top of each other.
Right panel: Functions ${\cal J}$ (dashed curve) and ${\cal I}$ (solid curve) as defined in (\ref{eq:J}) and (\ref{eq:I}), respectively.
\label{fig.3}}
\end{figure}

It is instructive to study some special limiting cases of the dispersion relations, where the mass dependence
of the thermal mass part of the excitation energies can be given explicitly by analytical formulae.
The gluon plasma frequency is defined as the long wavelength limit of the gluon dispersion relation
$\displaystyle \omega_{pl} = \lim_{k \to 0} \omega_{T,L} (k)$,
yielding after some calculations
\begin{eqnarray}
\omega_{pl}^2 &=& \frac{1}{9}g^2T^2 \label{eq:gluon_plasma_frequency}
\left(N_c + C_2 \sum_q^{N_f} {\cal J} \left( \frac{m_q}{T} \right)\right)\,\chi(g,\{m_q\}),\\
{\cal J}(x) &=& \frac{18}{\pi^2} x^2
\int\limits_0^\infty d\xi \, \frac{\xi^2}{\sqrt{1+\xi^2}} \label{eq:J}
\left( 1- \frac{\xi^2}{3(1+\xi^2)}\right) \frac{1}{1 + \exp ({x \sqrt{1 + \xi^2}})}
\end{eqnarray}
with the same value for the transverse and longitudinal polarizations, giving rise to a mass gap,
i.e. $\omega (k \to 0) > 0$. In general, $\chi(g,\{m_q\})$ depends on a vector $\{m_q\}$ with components $m_q$.
The function $\chi(g,m) = \chi(g,\{m,m\})$, depicted in the left panel of figure \ref{fig.3} for one independent mass $m=m_q$ and $N_f=2$, is implicitly defined by (\ref{eq:gluon_plasma_frequency}) with the numerically determined value of $\omega_{pl} = \omega(k\to0)$ on the l.h.s.
It approaches unity for small values of $g$ in agreement with the HTL result. At 
large values of the coupling $g$, $\chi(g,m)$ rapidly drops, making the plasma frequency soft.
In addition, $\chi(g,m)$ is almost independent of $m$ in the range of interest.
From this follows that the mass dependence of the plasma frequency is essentially contained
in the function ${\cal J}(m/T)$, which is depicted in figure \ref{fig.3}, right panel.
For large values of $m/T$,
${\cal J}(m/T)$ behaves like $\exp(-m/T)$
indicating the decoupling of the heavy quark sector from the thermal bath,
while for decreasing values of $m/T$, ${\cal J}(m/T)$ approaches unity with 
${m/T\to 0}$. Thus, in the limit $g \to 0$ and $m \to 0$ or $T\to \infty$, equation (\ref{eq:gluon_plasma_frequency})
gives the usual HTL gluon plasma frequency $\hat \omega_{pl}^2 = \frac{1}{9}g^2T^2(N_c+ C_2N_f)$.

In the asymptotic region, $k \rightarrow \infty$, the transverse gluon
dispersion relation is given by the gauge invariant result
$\omega^2=k^2+m_\infty^2$ with the asymptotic gluon mass
\begin{eqnarray}
m_\infty^2 &=&  \Re \Pi_T^g(k,k) = \frac16 g^2 T^2
\left(N_c + C_2 \sum_q^{N_f} {\cal I} \left( \frac{m_q}{T} \right)\right) ,
\label{eq:asy_gluon_mass} \\
{\cal I} (x) &=& \frac{12}{\pi^2} x^2 \int \limits_0^\infty d\xi \,
\frac{\xi^2}{\sqrt{1 + \xi^2}} \frac{1}{1 + \exp({x \sqrt{1 + \xi^2}})} \label{eq:I}.
\end{eqnarray}
The function $\cal I$ is depicted in the right panel of figure \ref{fig.3}.
The power expansion for small values of $m/T$ (cf.~Appendix A) reads
\begin{equation}
{\cal I} \left( \frac{m}{T}\right) = 1 + \alpha_2 \left( \frac{m}{T}\right)^2 +
\alpha_L \left( \frac{m}{T}\right)^2 \ln
\left( \frac{m}{T}\right)^2 + \alpha_4
\left( \frac{m}{T}\right)^4 +\sum_{j=3}^\infty
\alpha_{2j} \left( \frac{m}{T}\right)^{2j}
\label{eq:chiral_expansion}%(5.26)
\end{equation}
with coefficients $\alpha_n$ listed in table \ref{tab:chiral_expansion}
relegated to Appendix A. This expansion of $\cal I$ is convergent in the region $m/T < \pi$.
Note the term $\propto \alpha_L$ resembling chiral logarithms. 
Equations (\ref{eq:asy_gluon_mass}) and (\ref{eq:I}) highlight an important feature occurring
in the gluon sector, here, for the example of asymptotic dispersion relations.
As the quark
loop contribution is added to the gluon and ghost loops, the mass dependence
is not so striking in the range of $m$ we are interested in.
For instance, for $N_f = 2$ the term in brackets in (\ref{eq:asy_gluon_mass}) varies in the range
between $4$ and $3$ while going from $m=0$ to $m\to \infty$, evidencing the dominance of the strong non-Abelian self-coupling compared to the quark loop contributions in the thermal masses. In a QED plasma there is, of course, a much stronger dependence of the thermal photon mass on the electron mass $m_e$.

The heavy-quark expansion for $m/T \gg 1$ yields, as evident from (\ref{eq:chiral_expansion_bessel})
in Appendix A,
\begin{equation}
 {\cal I}\left( \frac{m}{T}\right) \simeq \frac{12}{\pi^2} \sqrt{\frac{\pi m}{2T}} \rme^{-\frac{m}{T}}.
\end{equation}
As ${\cal I} (m/T)$ becomes exponentially small for large values of $m/T$, the heavy quarks
decouple also in the asymptotic momentum region from the thermal bath (i.e., they are exponentially suppressed). 

\section{One-loop quark excitations \label{sec.22}}

\subsection{Quark self-energy}

The one-loop quark self-energy represented in the imaginary time formalism reads
\begin{eqnarray}
\Sigma(K) & = & - T \sum_n \int \frac{d^3p}{(2\pi)^3}
\big[ \Gamma^q_\mu {\cal S}(Q) \Gamma_\nu^q  {\cal D}_{\mu\nu} \big]
\label{eq:def:quark_selfenergy_1loop}
\end{eqnarray}
with loop momenta $P$ and $Q = K - P$.
The general structure of the self-energy reads
\begin{equation}
 \Sigma = a \, \feyndagg{K} + b\, \feyndagg{u} - c 
    =  \gamma_0\tilde b(\omega,k) -   {\boldsymbol \gamma}{\bf k}a(\omega,k) - c(\omega,k)  
 \label{eq:quark_selfenergy_general}
\end{equation}
with the medium 4-velocity $u_\mu$ and $\tilde b = \omega a+b$. 
After performing the Matsubara sum and continuing into Minkowski space-time one finds for the three independent self-energy functions
\begin{eqnarray}
\fl a(\omega,k) &=& \frac{g^2C_F}{2\pi^2 k^2} \int \limits_0^\infty d p \,
\left[\frac{p^2}{\E{p}}n_F(\E{p})
\left( 1 + \frac{h_F}{8 k p}\ln (a_F^+a_F^-) + \frac{d_F}{8 k p} \ln \frac{a_F^+}{a_F^-}\right)
\right. \nonumber \\
\fl & & + p n_B(p) \left. \left( 1 + \frac{h_B}{8k p}\ln (a_B^+a_B^-) + \frac{d_B}{8k p}
\ln \frac{a_B^+}{a_B^-}
- \frac{k}{4p} \ln (a_B^+a_B^-) \right) \right],
\label{eq:quarkSE_1loop_first}\\
\fl \tilde b(\omega,k) &=& \frac{g^2 C_F}{8\pi^2 k} \int \limits_0^\infty d p \, p
\left[n_F(\E{p})\ln \frac{a_F^+}{a_F^-} +n_B(p) \left(\ln \frac{a_B^+}{a_B^-}
- \frac{\omega}{p} \ln (a_B^+a_B^-)\right)\right], \\
\fl c(\omega,k) & = & m\frac{g^2C_F}{4\pi^2 k} \int \limits_0^\infty d p \, p
\left[ \frac{n_F(\E{p})}{\E{p}} \ln (a_F^+ a_F^-) - \frac{n_B(p)}{p}
\ln (a_B^+a_B^-)\right]  \label{eq:quarkSE_1loop_third}
\end{eqnarray}
with $C_F = (N_c^2 - 1)/(2 N_c)$, and the abbreviations read
\begin{eqnarray}
a_F^\pm &=& \frac{k^2-m^2-\omega^2 \pm 2\E{p}\omega-2pk}{k^2-m^2-\omega^2
\pm 2\E{p}\omega +2pk}, \\
a_B^\pm &=& \frac{k^2+m^2-\omega^2 \pm 2p\omega -2pk}{k^2+m^2-\omega^2
\pm 2p \omega + 2pk}, \\[1mm]
h_F &=& k^2-m^2-\omega^2, \quad d_F = 2 \E{p} \omega, \\
h_B &=& k^2+m^2-\omega^2, \quad d_B = 2 p \omega \, .
\label{eq:quarkSE_1loop_last}
\end{eqnarray}

\subsection{Quark dispersion relations}

The dispersion relations are determined by the poles of the 
resummed quark propagator ${\cal S}$,
given by the Dyson equation ${\cal S}^{-1} = {\cal S}_0^{-1} + \Sigma$, 
leading to
\begin{equation} 
{\cal S}^{-1} = \gamma_0(\omega+ \tilde b) - 
{\boldsymbol \gamma }{\bf k} (1+ a) - (m + c).
\end{equation}
With the definitions of projectors relegated to Appendix B, this can be rewritten yielding
\begin{equation}
{\cal S} = \frac{{\mathfrak P}^+_{{\bf k},m}\gamma_0}{r(n\omega-{\cal E})}
     + \frac{{\mathfrak P}^-_{{\bf k},m}\gamma_0}{r(n\omega+{\cal E})}.
\label{eq:quark_propagator_decomposed}
\end{equation}
The quark (plasmino) dispersion relation is obtained as the positive energy solution of
$n\omega_q -{\cal E} = 0$ ($n\omega_p +{\cal E} = 0$).
The dispersion relations for the quasi-particles may be parametrized as
$\omega_q^2 = k^2 +m^2 + m_f^2 {\cal F}_q$ and $\omega_p^2 = k^2 + m_f^2 {\cal F}_p$
for quarks and plasminos, respectively.
The functions ${\cal F}_{p,q}$,which non-trivially depend on $k$ and $m$, encode the effects on the excitations induced by the thermal medium 
with temperature $T$.
For fixed $k$ and $m$ these functions multiplied by $m_f^2$ serve as effective thermal mass parameters. 
${\cal F}_q = 0$ accounts for a simple on-shell dispersion relation, while ${\cal F}_p = 0$ represents the light cone.
The quantity $m_f^2 = \frac{1}{8}C_F g^2 T^2$
is the usual HTL quark
plasma frequency, setting a typical energy scale in these studies~\cite{LeBellac}.

\begin{figure}[ht]
% \psfrag{g = 0.3}{\tiny $g=0.3$}
% \psfrag{g = 1.0}{\tiny $g=1.0$}
% \psfrag{g = 3.0}{\tiny $g=3.0$}
% \psfrag{x}{\small $k/m_f$}
% \psfrag{y}{\small $m/T$}
% \psfrag{Z}{\small \hspace*{-2.5mm} ${\cal F}_q$}
% \psfrag{Z2}{\hspace*{6cm} \small ${\cal F}_p$}
\begin{center}
\includegraphics[scale=0.65]{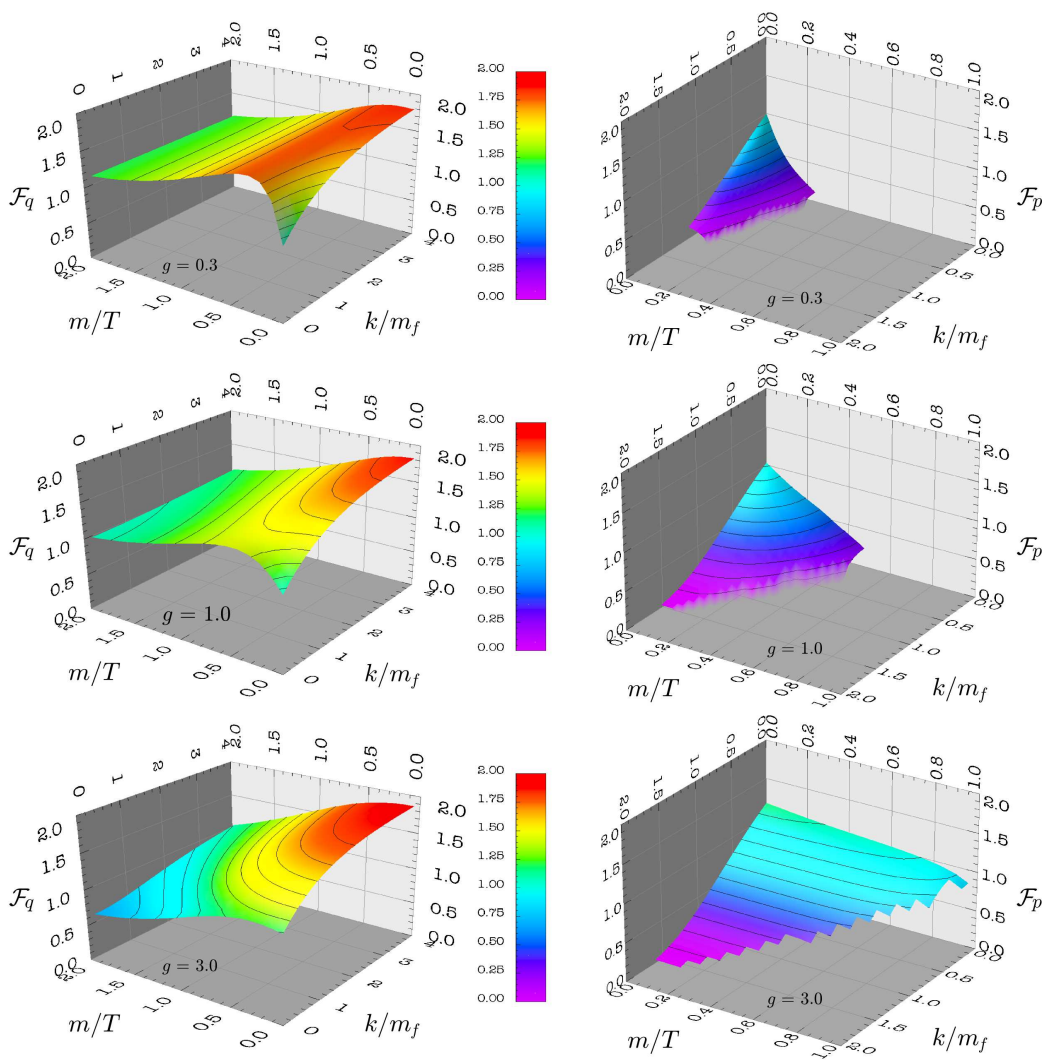}
\end{center}
\caption{Quark (left) and plasmino (right) dispersion relations represented as ${\cal F}_q = (\omega_q^2-k^2-m^2)/m_f^2$ for regular quark excitations and ${\cal F}_p = ({\omega^2_p -k^2})/m_f^2$ for plasmino excitations as functions of scaled momentum and scaled quark mass
for $g = 0.3, \, 1.0, \, 3.0$ from top to bottom.
For a better representation, the regular quark and the plasmino quantities are considered from different perspectives. Faint isolines are for ${\cal F}_{q,p} = 0.1,0.2, \ldots 1.9$.
The representation of ${\cal F}_p$ in the right panel ends where numerically no real solutions of the dispersion
relations can be found.
\label{fig.4}}
\end{figure}

A survey of this representation of the dispersion relations is exhibited in figure \ref{fig.4},
where the temperature dependent parts ${\cal F}_q$ and ${\cal F}_p$ for regular quark and collective plasmino excitations are shown in the left and right columns, respectively.
Again, different cuts allowing for a better quantitative representation are relegated to Appendix C (cf.~figures \ref{fig.7} and \ref{fig.8}).
Further examinations of the one-loop self-energies show 
that these exhibit non-zero imaginary parts, both, below and above the light cone 
which are particularly large below the light cone.

Considering, instead, $\omega_p^2 = k^2 +m_f^2{\cal F}_p$ as a function of $k^2$,
the plasmino branch exhibits a local minimum which, however, vanishes for $m > gT$.
This local minimum gives rise to van Hove singularities in certain emission rates \cite{Thoma_Peshier}.

In addition, for any value of $m > 0$, the collective branch completely disappears from the spectrum for momenta larger than a certain critical one.
Furthermore, the plasmino branch lies energetically below the regular quark
excitation branch $\omega_q^2$ (Surprisingly, the lattice QCD calculations \cite{Karsch-Kitazawa}
show an opposite order at $k=0$).

As mentioned above, with some replacements (here, one has to use $C_F \to 1$ in addition) 
the electron excitations in
an electron-positron QED plasma are also described. Particularly interesting would
be an experimental verification of the purely collective plasmino excitations.

\subsection{Quark plasma frequency and asymptotic behaviour} 

For vanishing momenta $k \to 0$, the plasma frequencies for quarks ($+$) and plasminos ($-$)
$\omega_\pm = \omega(k\to 0)$ are implicitly given by
\begin{equation}
\omega_\pm \mp m + \left. \Big( \tilde b \mp c \Big) \right|_{k\to0}  =  0,
\end{equation}
where we used $ \left. (k a)\right|_{k\to0} = 0$ due to rotational invariance.
The long-wavelength limit of the quark dispersion relation is therefore independent of the function $a$.
An explicit calculation of the various limits yields the implicit equation
\begin{eqnarray}
 0 &=&\omega_\pm \mp m + \frac{g^2C_F}{2\pi^2} \int \limits_0^\infty \frac{d p \, p}{z_\pm^2-p^2}
    \left[
        n_F(\epsilon_{\bf p}) \frac{p\epsilon_{\bf p}}{\omega_\pm^2} (\omega_\pm \mp m)\right. \nonumber \\
     & &\hspace*{4cm} \left. + \frac{n_B(p)}{\omega_\pm} \Big( (p^2\mp m z_\pm) + z_\pm(\omega_\pm \mp m) \Big)
    \right]\,  \label{eq:quark_plasmafreq_implicit}
\end{eqnarray}
with $z_\pm = (m^2-\omega_\pm^2) / 2 \omega_\pm$. 
The splitting of the two branches is of order $m$; in the chiral limit, $m \to 0$, both branches meet at $k = 0$. 
In the limit $g\ll 1$ and $m \lesssim gT$, one recovers a formula given 
by Pisarski \cite{Pisarski}
\begin{equation}
\omega_\pm = \frac{1}{2} \left( \sqrt{m^2+4m_f^2} \pm m \right),
\end{equation}
and the difference of the plasma frequencies of quarks and plasminos becomes $\omega_+ - \, \omega_- = m$.
Thus, the regular quark has a higher excitation energy than the plasmino. In
the limit $m=0$ the high temperature result $\omega_\pm = m_f$ is obtained.

In the asymptotic region, the momentum $k$ is the largest scale, allowing for some approximations
to obtain an explicit analytic expression for the asymptotic dispersion relation of the regular quark excitation.
In this case, the equation $n\omega - {\cal E} = 0$ can be transformed to yield
\begin{eqnarray}
 \omega^2 = k^2 + m^2 + 2 \left.\Big( k^2 \, a + k \, \tilde b \Big)\right|_{k\to \infty},
\end{eqnarray}
which is completely independent of the self-energy function $c$.
The asymptotic quark dispersion relation can then be written as
\begin{equation}
\omega^2=k^2+m^2 + 2M_+^2, \quad 
M_+^2= \frac{1}{3} m_f^2 
\left({\cal I}\left(\frac{m}{T}\right) + 2\right)
\label{M_plus}
\end{equation}
with the integral $\cal I$ given in (\ref{eq:I}) 
and the same expansion as in (\ref{eq:chiral_expansion}).
For $m \to 0$, one finds $M_+^2 \to m_f^2$. 

\section{Gauge dependence \label{sec.3}}

As the HTL approximation has been proved to be gauge independent
\cite{Kunstatter-Kobes-Rebhan}, one may look for regions in parameter
space where the self-energies in one-loop approximation coincide
with HTL results. 
The real parts of the HTL and one-loop self-energies coincide for small values 
of $\omega$ and $k$, of course, but for larger
values of $\omega$ and/or k both may deviate noticeably in general. 
However, near the light cone, i.e. for $\omega \approx k$, they are approximately equal 
, and on the light cone, $\omega = k$, HTL and one-loop self-energies coincide,
cf.~\cite{Bluhm-EPJC,Seipt-diploma}. 
The self-energies slightly above the light-cone determine the excitation energies of the quasi-particles
in a wide parameter range and are, therefore,
relevant for the quasi-particle model \cite{Peshier,Bluhm-PLB}.

The gauge independence of the asymptotic gluon mass $m_\infty^2$ can be shown by noting that the transverse parts of the Yang-Mills contributions to the gluon self-energy at the light cone $\Pi_T^{YM}(k,k)$ are exactly the same in the one-loop and HTL approximations. Furthermore, the quark contribution $\Pi_T^q$ is inherently gauge independent at one-loop level. Hence, the asymptotic thermal gluon mass is a gauge invariant quantity.

To check further for the gauge dependence of the above one-loop results in
Feynman gauge -- here in particular we focus on the asymptotic mass of regular quarks -- we compare them with explicit calculations in Coulomb gauge. For large momenta,
the regular quark excitation energies in Feynman gauge ($\omega$) and Coulomb gauge ($\omega_C$) are related via
\begin{eqnarray}
\omega_C^2(\omega, k) &=& \omega^2 (\omega, k) +\Delta_C(\omega,k), \\
  \Delta_C(\omega,k) & = & \frac{g^2C_F}{\pi^2}
\int \limits_0^\infty d p \, \frac{p^2}{\E{p}} n_F(\E{p})
\left( 1 - \frac{k^2+p^2}{4k p} \,
\ln \frac{k^2+p^2+2k p}{k^2+p^2-2k p} \right), 
\end{eqnarray}
with $\displaystyle \lim_{k\to\infty} \Delta_C(\omega ,k)\to 0$, thus proving
the coincidence of the dispersion relations in both gauges in the asymptotic region.

\section{Summary \label{sec.4}}

In summary, we survey the quark mass dependence of thermal one-loop
self-energies and emerging dispersion relations of quarks and gluons
in QCD in Feynman gauge. The motivation and the related focus of our
presentation are given by upcoming lattice QCD results on spectral
properties of quarks and gluons in a hot and deconfined medium and
the need of chiral extrapolations in bulk properties of the
quark-gluon plasma. While the results of \cite{Karsch-Kitazawa} are
at variance with the energetic ordering of quark and plasmino
excitations from perturbative QCD, they otherwise confirm the existence of a mass gap of
quarks in line with basic assumptions of the successful
phenomenological quasi-particle model \cite{Peshier,Bluhm-PLB}. The
present analysis challenges the chiral extrapolation of the equation
of state performed in \cite{Bluhm-POS}, as the important term $\propto mM_+$ in
the quark excitation dispersion relation is not supported by our
one-loop results. This may be considered as further hint to
strong non-perturbative effects in the quark-gluon plasma in a range
accessible in present and future heavy-ion collisions.
Indeed, the asymptotic Dyson-Schwinger type approach outlined in Appendix D,
points to severe corrections to the one-loop results in the strong coupling regime
for fermion masses of the order of the temperature.
Progressing
lattice QCD results are required to reveal the fundamental
excitations of deconfined matter.

\section*{Acknowledgements}

The authors thank Frithjof~Karsch and Munshi~G.~Mustafa for valuable discussions.
The work is supported by BMBF 06DR136 and EU-I3HP.

\appendix

\section{Expansion of the function ${\mathcal I}$}

To derive the expansion in (\ref{eq:chiral_expansion}) of the function ${\cal I}$ given in (\ref{eq:I}) we rewrite the expression using
$(\E{p} + 1)^{-1} = \sum_{j=1}^\infty (-1)^{j+1} \rme^{-j\E{p}}$
to
\begin{equation}
{\cal I}(x) = \frac{12}{\pi^2T^2}\sum_{j=1}^\infty (-1)^{j+1} \int \limits_0^\infty d p
\frac{p^2}{\E{p}} \rme^{-j\E{p}}
\label{eq:chiral_expansion_integral}
\end{equation}
and obtain
\begin{equation}
{\cal I} (x) = \frac{12}{\pi^2} \sum_{j=1}^\infty (-1)^{j+1} \frac{x}{j} K_1(j x)
\label{eq:chiral_expansion_bessel}
\end{equation}
with $x = m/T$. For small quark masses, $x \to 0$, one may represent $\cal I$ as
\begin{equation}
{\cal I} (x) = \frac{1}{2\pi i}\frac{12}{\pi^2} \sum_{j=1}^\infty \int
\limits_{c-i\infty}^{c+i\infty} d s\,
(-1)^{j+1} j^{-s-1} \, x^{1-s} \,{\cal M}[K_1; s] 
\label{eq:chiral_expansion_z3}
\end{equation}
with the Mellin transform
\begin{eqnarray}
{\cal M}[K_1;s] & = & \int \limits_0^\infty dz \, z^{s-1} K_1(z) =
2^{s-2} \, \Gamma \Big(\frac{s-1}{2} \Big)
\, \Gamma \Big(\frac{s+1}{2} \Big)
\end{eqnarray}
of the first modified Bessel function of the second kind $K_1$.\\
Exploiting $\sum_{j = 1}^\infty (-1)^{j+1} j^{-s-1} = \left( 1 - 2^{-s} \right) \zeta(s+1),$
with the Riemann zeta function $\zeta$, we find as intermediate step
\begin{equation}
{\cal I} (x) = \frac{12}{\pi^2}\frac{1}{2\pi i}
\int \limits_{c-i\infty}^{c+i\infty} d s \,\frac{x^{1-s}}{2^{2-s}}
\left( 1-\frac{1}{2^s}\right) \zeta(s+1)
\, \Gamma \Big(\frac{s-1}{2} \Big)
\, \Gamma \Big(\frac{s+1}{2} \Big) ,
\label{eq:chiral_expansion_sum_done}
\end{equation}
which can be evaluated with a suitable contour and appropriate $c$
\cite{Seipt-diploma} to yield
\begin{eqnarray}
{\cal I} &=& \sum_i \left.\mbox{Res}\, f(s)\right|_{s=s_i}, \\
f(s) &=& \frac{6}{\pi^2}\, \frac{x^{1-s}}{2^{1-s}} \left( 1-\frac{1}{2^s}\right) \zeta(s+1)
\, \Gamma \Big(\frac{s-1}{2} \Big) \, \Gamma \Big(\frac{s+1}{2} \Big).
\end{eqnarray}
Further evaluation leads to the series in (\ref{eq:chiral_expansion})
with coefficients given in table \ref{tab:chiral_expansion}.
Although the numerical value of the coefficients rapidly approaches zero, this expansion has a finite radius of convergence, which is 
\begin{equation} 
R^2 = \lim_{n\to \infty} \left| \frac{\alpha_n}
{\alpha_{n+2}}\right| = \pi^2, \qquad \, n\geq 4 \, .
\end{equation}

\begin{table}[htp]
\caption{The coefficients of the chiral expansion in (\ref{eq:chiral_expansion}).
$\gamma_E$ is the Euler-Mascheroni number and $\zeta$ stands for Riemann's zeta function.}
\begin{indented}
\item[]\begin{tabular}{lll}
\br
coefficient & analytic expression & numerical value \\
\mr
$\alpha_2$ & $ -\frac{3}{\pi^2}(\ln \pi + \frac{1}{2} - \gamma_E)$ & $\approx -0.32449$ \\[4pt]
$\alpha_L$ & $ \frac{3}{2\pi^2} $ & $ \lineup\approx \m 0.15198 $ \\[4pt]
$\alpha_4$ & $-\frac{21}{16\pi^4} \zeta(3)$ & $\approx -1.619 \cdot 10^{-2} $ \\[4pt]
$\alpha_6$ & $\frac{93}{128\pi^6} \zeta(5) $ & $ \lineup\approx \m 7.836 \cdot 10^{-4} $ \\[4pt]
$\alpha_8$ & $ -\frac{1905}{4096 \pi^8} \zeta(7)$ & $\approx -4.943 \cdot 10^{-5} $ \\[4pt] \mr
$\alpha_{n}$ & $ (-1)^{\frac{n-2}{2}} \, \frac{24}{n} \, \frac{2^{n-1}-1}{(2\pi)^n}
        \, \frac{(n-3)!!}{(n-2)!!} \zeta(n-1) $ &  \\ 
 & \hspace*{2.5cm} for $n = 4,6,8,\ldots$ & \\ \br
\end{tabular}
\end{indented}
\label{tab:chiral_expansion}
\end{table}

\section{Quark propagator}

The explicit breaking of chiral symmetry drastically modifies the structure of the quark propagator compared
to the usually treated case, where all current quark masses are set to zero. As a consequence, the usual projectors, which allow for a distinction of the different quasi-particle states, are not applicable.
Thus, one has to examine in some detail the spinor structure 
of the propagator, which provides projectors 
to the relevant subspaces of the physical (quasi-particle) excitations.
For the self-energy we employ the decomposition
given in (\ref{eq:quark_selfenergy_general}).
Utilizing Dyson's equation 
yields the general structure for the inverse quark propagator
\begin{equation}
 {\cal S}^{-1} = 
 r \left( n \omega \gamma_0 - {\boldsymbol\gamma}{\bf k} - {\cal M} \right) 
 \label{eq:quark_propagator_general}
\end{equation}
with the three self-energy functions 
$r \equiv (1+a)$, $n \equiv 1+b/\omega r $ and ${\cal M} \equiv (m+c)/(1+a)$, 
depending on $\omega, |{\bf k}|$ and $m$. 
Although there is an explicit breaking of chiral symmetry due to the finite current quark masses, chiral symmetry is assumed not to be broken spontaneously at high temperature,
thus, $\displaystyle \lim_{m \to 0} {\cal M}(\omega,k,m) = 0$.
In contrast to the vacuum, an additional function $n$ appears, 
which may be interpreted as an analogue to the refraction index in optics \cite{WeldonHole}. 
The appearance of this function is due to the breaking of Lorentz invariance 
by the term $\propto \feyndagg{u}$ in (\ref{eq:quark_selfenergy_general}).

The spinors $\psi$ representing the quasi-particle states
are supposed to be solutions of the Dirac equation
\begin{equation}
(n\omega\gamma_0 - {\boldsymbol \gamma}{\bf k} - {\cal M}) \psi(k) = 0.
\label{hamiltonian}
\end{equation}
Non-trivial solutions are found only for $n^2\omega^2 = {\cal E}^2$  
with ${\cal E} = +\sqrt{k^2+{\cal M}^2}$. 
This is an implicit equation for the excitation energies $\omega$ of the system, 
leading to $\omega = \pm {\cal E} / n$.
A Hamiltonian may be defined by
\begin{equation}
{\cal H} = \frac{1}{n}(\gamma_0 {\boldsymbol \gamma} {\bf k} +\gamma_0 {\cal M}) 
\end{equation}
allowing for rewriting (\ref{hamiltonian}) as 
${\cal H} \psi(k) = \omega \psi(k)$. 
Let $\omega_a$ be a solution of this equation, 
then we have $n_a \omega_a = \pm {\cal E}_a = \pm \sqrt{k^2+{\cal M}_a^2}$, 
where $n_a=n(\omega_a,k,m)$ and ${\cal M}_a = {\cal M}(\omega_a,k,m)$,
and the states can be classified by the eigenvalues of the sign operator
$\displaystyle \Lambda_{\cal H} = n {\cal H}/ {\cal E}$, 
which obviously commutes with $\cal H$ and has eigenvalues $\lambda_{\cal H} = \mbox{sgn}(n\omega) = \pm 1$.

In the limit $m\to0$, the operator $\Lambda_{\cal H}$ 
becomes the operator of chirality times helicity,
$\Lambda = \gamma_0 {\boldsymbol \gamma}{ \bf \hat k}$,
where ${\bf \hat k} = {\bf k} / |{\bf k}|$.
Thus, the states with eigenvalue $+1$ have the same value for chirality and helicity, 
whereas the states with eigenvalue $-1$ have the opposite value for chirality and helicity. 
This simple interpretation fails for $m \neq 0$, 
as chirality is not a conserved quantum number in this case.

The projectors that decompose the Dirac structures according to the physical 
(quasi-particle or collective excitation) states are induced by the operator $\Lambda_{\cal H}$ and read
\begin{equation} 
{\mathfrak P}^\pm_{ {\bf k},m} = \frac{1}{2}(1\pm \Lambda_{\cal H}) =\frac{1}{2}\left[1 \pm
 \gamma_0\frac{ {\boldsymbol\gamma}{\bf k}  + {\cal M} } { {\cal E}}\right],
\end{equation}
satisfying 
${\mathfrak P}^\pm_{{\bf k},m}{\mathfrak P}^\pm_{{\bf k},m} = {\mathfrak P}^\pm_{{\bf k},m}$
and ${\mathfrak P}^+_{{\bf k},m} {\mathfrak P}^-_{{\bf k},m} = 0$.
Then, ${\mathfrak P }^+_{ {\bf k},m}$ (${\mathfrak P}^-_{{\bf k},m}$) 
projects on all possible states with $\mbox{sgn}(n\omega)>0$ ($\mbox{sgn}(n\omega)<0$).
In the chiral limit $m \to 0$, they take the form 
${\mathfrak P}^\pm_{\bf k} = \frac{1}{2}(1\pm\gamma_0{\boldsymbol \gamma}{\bf \hat k})$, 
and for $k\to 0$ one obtains ${\mathfrak P}^\pm_{m} = \frac{1}{2}(1\pm\gamma_0)$, which are often employed decompositions. Our more generally decomposed quark-propagator
for $m \geq 0$ takes eventually the form of (\ref{eq:quark_propagator_decomposed}).

\section{Cuts through the dispersion relations}

\begin{figure}[t]
\include{general_frag}
\psfrag{ylabel}{${\cal G}_{T}$}
\psfrag{y2label}{${\cal G}_{L}$}
\psfrag{xlabel}{$m/T$}
\psfrag{smallest}{$k/gT = 0$}
\psfrag{center}{$k/gT = 1$}
\psfrag{largest}{$k/gT = 10$}
\psfrag{asymptotic}{$m_\infty^2$}
\includegraphics[scale=0.57]{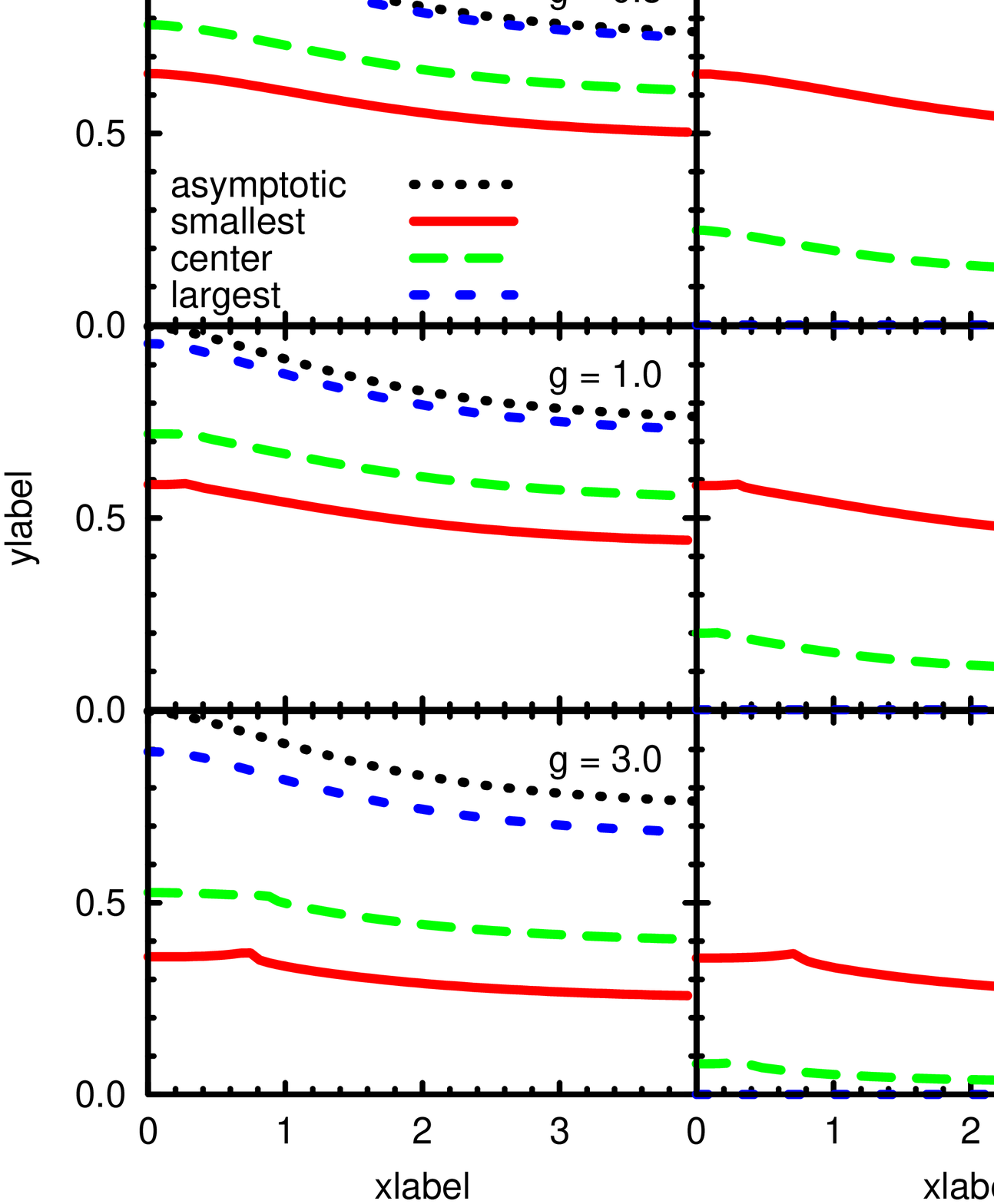}
\caption{Cuts through the dispersion relations exhibited in figure \ref{fig.2}
as a function of $m/T$ for $k / g T = 0$ (red solid curves),
$1$ (green long-dashed curves), $10$ (blue short-dashed curves) and the asymptotic
region ($k \to \infty$, black dotted curves)  for 
$g = 0.3, \, 1.0, \, 3.0$ from top to bottom. 
Left: Transverse gluon branch ${\cal G}_T$, right: Longitudinal gluon branch ${\cal G}_L$. 
This figure exposes the dependence on $m$.
\label{fig.5}}
\end{figure}

\begin{figure}[t]
\include{general_frag}
\psfrag{ylabel}{${\cal G}_{T}$}
\psfrag{y2label}{${\cal G}_{L}$}
\psfrag{xlabel}{$k/gT$}
\psfrag{smallest}{$m/T = 0.0$}
\psfrag{center}{$m/T = 0.1$}
\psfrag{largest}{$m/T = 1.0$}
\includegraphics[scale=0.57]{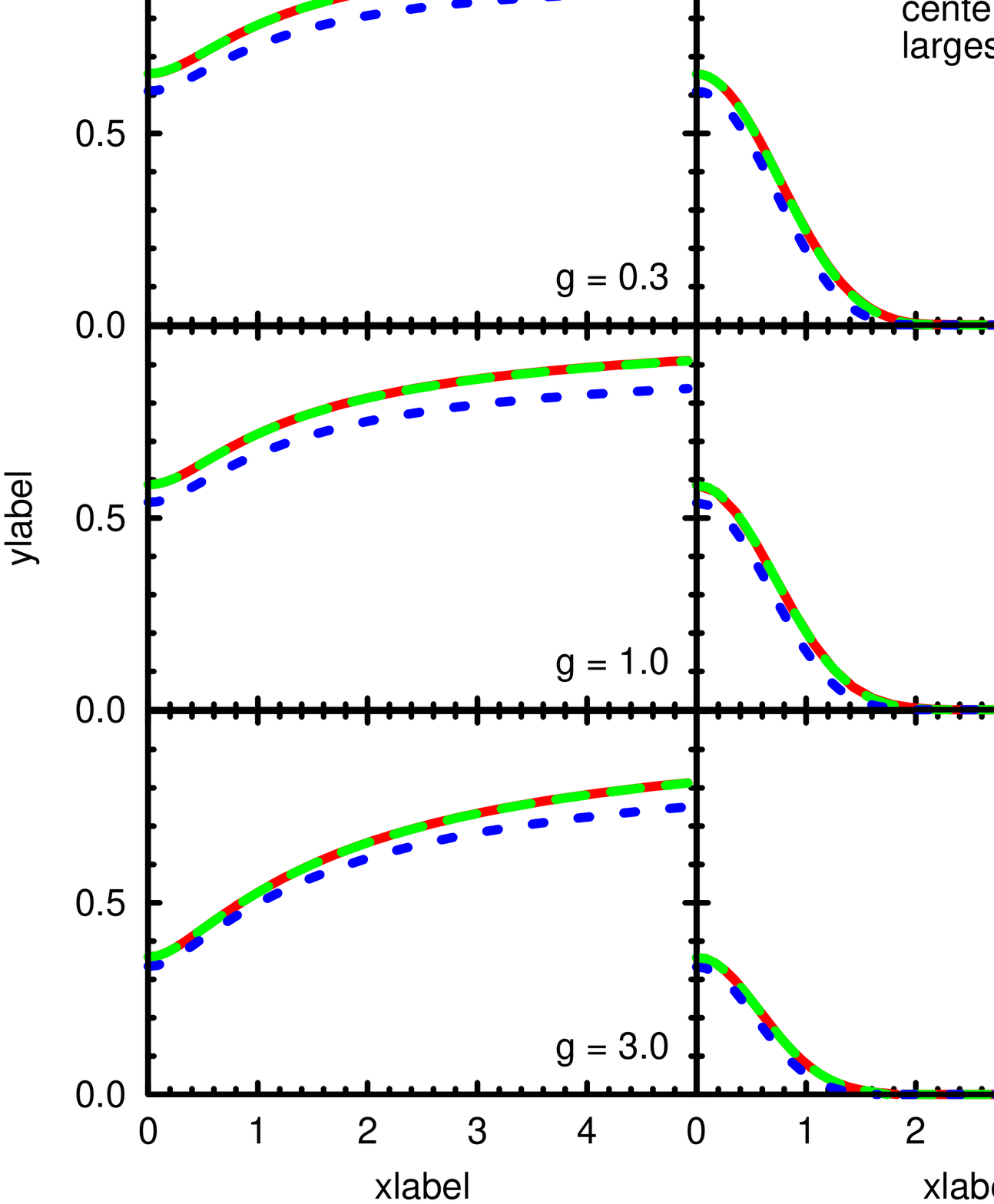}
\caption{Cuts through the dispersion relation exhibited in figure \ref{fig.2} for $m/T = 0$ (red solid curves), $0.1$ (green long-dashed curves) and $1$ (blue short-dashed curves) 
and for $g = 0.3, \, 1.0, \, 3.0$ from top to bottom. 
Left: Transverse gluon branch, right: Longitudinal gluon branch. 
This figure exposes the momentum dependence.
\label{fig.6}}
\end{figure}

We present here various cuts of $m = const$, $g = const$ and $k = const$
to quantify the dispersion relations and expose the relevant dependencies.
Figure~\ref{fig.5} exhibits ${\cal G}_{T,L}$ from figure \ref{fig.2} in section \ref{sec.21}.B
as a function of $m/T$ for various fixed values
of momentum $k$ (different curves) and couplings $g$ (different panels).
The mild mass dependence mentioned in section~\ref{sec.21}.B is clearly seen in this representation.
For longitudinal modes, $\omega(k)$ approaches the light cone for larger momenta.
In addition, at small $k$ the dependence on the coupling $g$ is strong for all
considered values of $m$, while at large momenta the coupling strength is of
minor importance (cf. transverse modes).
Note the kinky structure for small values of $k$.
The momentum dependence of ${\cal G}_{T,L}$ is displayed in figure \ref{fig.6} for various
fixed values of $m/T$ (different curves) and various fixed values of the
coupling $g$ (different panels). Again, the weak $m$ dependence is clearly visible.
Note also the fairly strong dependence on the coupling strength.

\begin{figure}[t]
\include{general_frag}
\psfrag{ylabel}{${\cal F}_{q}$}
\psfrag{y2label}{${\cal F}_{p}$}
\psfrag{xlabel}{$m/T$}
\psfrag{smallest}{$k/m_f = 0$}
\psfrag{center}{$k/m_f = 1$}
\psfrag{largest}{$k/m_f = 10$}
\psfrag{asymptotic}{$M_+^2$}
\includegraphics[scale=0.57]{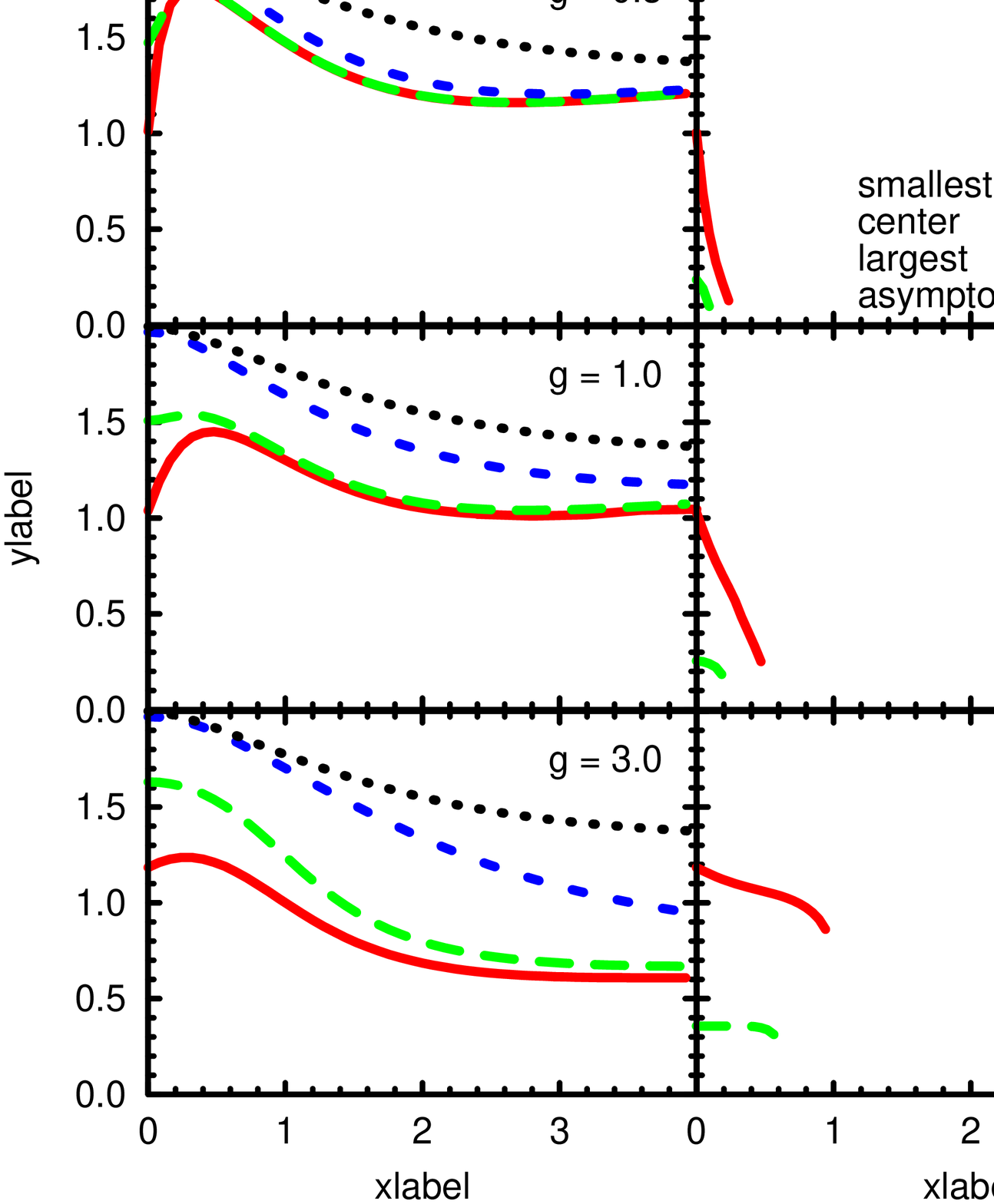}
\caption{Cuts through the dispersion relations in figure \ref{fig.4} as a function of the scaled quark mass for $k / m_f = 0$ (red solid curves), $1$ (green long-dashed curves), 
$10$ (blue short-dashed curves) and asymptotically large values ($M_+^2$, black dotted curves)
and for $g = 0.3, \, 1.0, \, 3.0$ from top to bottom. Left: Normal quark branch (${\cal F}_q$),
right: Plasmino branch (${\cal F}_p$). This figure exposes the dependence on $m$.
\label{fig.7}}
\end{figure}

\begin{figure}[t]
\include{general_frag}
\psfrag{ylabel}{${\cal F}_{q}$}
\psfrag{y2label}{${\cal F}_{p}$}
\psfrag{xlabel}{$k/m_f$}
\psfrag{smallest}{$m/T = 0.0$}
\psfrag{center}{$m/T = 0.1$}
\psfrag{largest}{$m/T = 1.0$}
\includegraphics[scale=0.57]{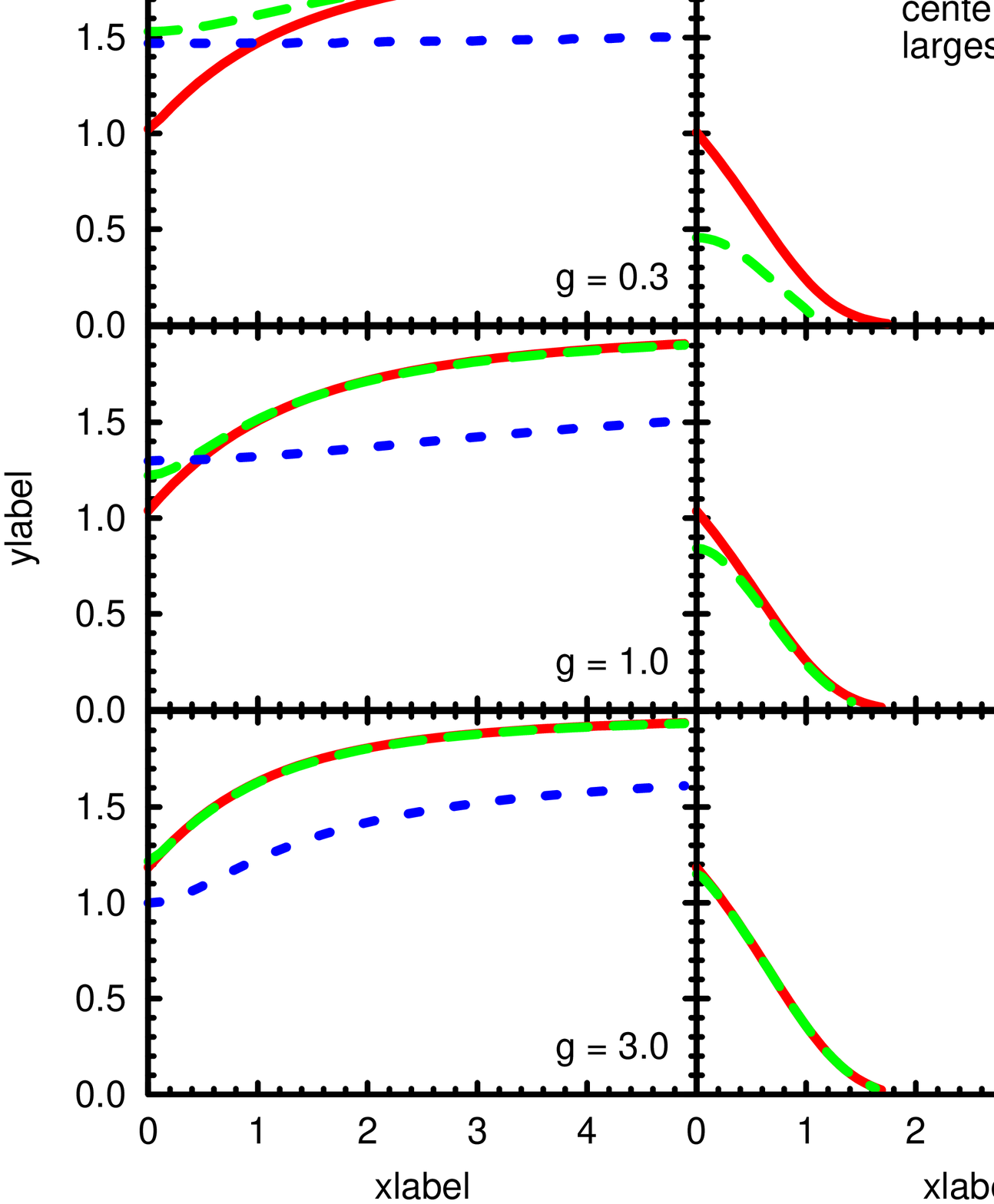}
\caption{Dispersion relations ${\cal F}_{q,p}$ vs.\ scaled $k$ as in figure \ref{fig.4} 
but for various cuts with
$m/T = 0$ (red solid curves), $0.1$ (green long-dashed curves) and $1$ (blue short-dashed curves) for $g=0.3, 1.0, 3.0$ from top to bottom. Left: Normal quark branch,
right: Plasmino branch. This figure exposes the momentum dependence.
\label{fig.8}}
\end{figure}

Figures \ref{fig.7} and \ref{fig.8} exhibit various cuts through the dispersion relations of quark excitations
shown in figure \ref{fig.4} in section~\ref{sec.22}.B in an analog manner as presented for the gluons above.
Figure \ref{fig.7} exposes the strong quark mass dependence for
various fixed values of momentum (different curves) and coupling strengths
(different panels). Interesting is the non-monotonic behaviour at small momenta
which is most pronounced for weak coupling. Striking is the disappearance
of the plasmino branch at large momenta but also for smaller momenta and increasing mass.
(see right panels). The plasmino branch (if there is any) persists for larger values of $m$
the larger the coupling $g$ is. The energy splitting with changing coupling is most severe for small momenta.

The momentum dependence of ${\cal F}_q$ and ${\cal F}_p$ is exhibited in 
figure \ref{fig.8} for various fixed
values of $m/T$ (different curves) and various values of the coupling $g$
(different panels). For normal quark excitations ${\cal F}_q$ rises monotonically 
with increasing momenta
(left panels), while for plasmino excitations ${\cal F}_p$ drops with increasing momenta (right panels).
For strong coupling $g$, the plasmino excitations become independent of the quark mass 
(right bottom panel), while in the weak coupling regime a severe mass dependence
is visible (right top panel). Note also the approximate mass independence
of ${\cal F}_q$ for small quark masses in the strong coupling regime (left bottom panel).

In the quasi-particle description of QCD thermodynamics (cf.~\cite{Peshier,Bluhm-PLB}), equation (\ref{eq:asy_gluon_mass}) is used with $\sum_q^{N_f}{\cal I} \to N_f$,
i.e. the quark mass dependence in the asymptotic gluon dispersion relation is disregarded.
Inspection of the left panels of figure \ref{fig.5} supports this approximation:
The energy of transverse gluon excitations (say, with momenta $k \sim g T$)
increases only by a tiny amount, in particular for larger $g$, when decreasing the quark mass from $m/T = 0.4$ (as used in \cite{Allton} for ''light quarks'') to the chiral limit $m = 0$. For larger values of $k$ this
increase is still less than 4 \%.
Curiously, in the chiral limit, the one-loop transverse gluon excitations obtain a larger thermal mass, thus, the according entropy carried by these modes is reduced.

Concerning the asymptotic (large momenta) regular quark dispersion relation, a bi-linear term $2m M_+$ is employed additionally in the quasi-particle model~\cite{Peshier,Bluhm-PLB} which is absent in (\ref{M_plus}).
(In fact, also any quark mass dependence in $M_+^2$ is disregarded in~\cite{Peshier,Bluhm-PLB}, thus, reducing $M_+^2$ to $m_f^2$.)
The term  $2m M_+$ used in~\cite{Peshier,Bluhm-PLB} was inspired by \cite{Pisarski}, mimicking the quark mass
dependence of regular quark excitations at small momenta and small $m$ and $g$ (see also red curves in the left panels of figure \ref{fig.7} for small $m$ and $g$). There, in particular for small and moderate couplings,
${\cal F}_q$ drops by a significant amount (about 50 \%) when decreasing $m$ from $m/T = 0.4$ to the chiral limit $m = 0$.
As a consequence, the entropy attributed to such excitations increases when lowering the quark mass.
In \cite{Bluhm-POS} (last reference) it was shown that with such an ansatz for the quasi-quark dispersion relation by adjusting the quasi-particle
model parameters to lattice QCD results~\cite{Allton} (first reference) using $m/T=0.4$ for ''light quarks''
good agreement with lattice
QCD thermodynamics \cite{cheng} for ''almost physical quark masses'' was found when extrapolating
to the corresponding quark mass values.

In contrast, based on the one-loop considerations presented in this work, the thermal mass contribution ${\cal F}_q$ of regular quark excitations
increases with decreasing $m$ for large momenta (${\cal F}_q = 2 M_+^2/m_f^2$ resembling the behaviour of the function $\cal I$ with $m$), as seen in the left panels of figure \ref{fig.7}.
If such a quark mass dependence (i.e. equation (\ref{M_plus})) would have been implemented in the quasi-particle model, the quark mass extrapolation reported in \cite{Bluhm-POS} would fail to reproduce the corresponding lattice QCD results \cite{cheng}.
This highlights the subtle role of the
actual dispersion relation employed in the quasi-particle model, as mentioned
already in \cite{susceptibilities} and points to the necessity of a term $\propto m M_+$ of sufficient
strength in order to account for the quark mass dependence in the non-perturbative regime.

It happens that for momenta $k/T \sim {\cal O}(1)$ and small $m$, the small-momentum ansatz $\alpha m M_+$, where $\alpha \approx 2$, represents a better approximation to the mass dependence of the regular quark dispersion
relation than the asymptotic form in (\ref{M_plus}) (see left panels of figure \ref{fig.7}). This motivates to some extent the dispersion relations employed in the model and
the performed chiral extrapolation in~\cite{Bluhm-POS}.
(Note that the quasi-particle model~\cite{Peshier,Bluhm-PLB} does not involve plasmino and plasmon excitations; their contributions to the entropy density, for instance, are found to be numerically small \cite{Schulze}.)

\section{Asymptotic Dyson-Schwinger approach in Abelian gauge theory}

In order to estimate higher-loop order correction effects on our one-loop results, we contrast the one-loop
results with corresponding calculations obtained in a Dyson-Schwinger type approach. In the following, we focus on
the fermion mass dependence in the asymptotic thermal mass expressions.
The Dyson-Schwinger equations (DSE) for the 2-point functions, here presented for an Abelian gauge theory, read
\begin{fmffile}{DSE}
\begin{eqnarray}
\left[
\parbox{30mm}
{
	\begin{fmfgraph*}(80,40)
	 \fmfleft{i}\fmfright{o}
	 \fmf{vanilla,tension=3.}{i,b,o}
	 \fmfblob{.12w}{b}
	\end{fmfgraph*}
}
\right]^{-1} &=&
\left[
\parbox{30mm}
{
	\begin{fmfgraph*}(80,40)
	 \fmfleft{i}\fmfright{o}
	 \fmf{vanilla,tension=3.}{i,o}
	\end{fmfgraph*}
}
\right]^{-1} +
\parbox{25mm}
{
	\begin{fmfgraph*}(80,45)
	 \fmfleft{i}\fmfright{o} \fmftop{bb}
	 \fmf{vanilla,tension=3.}{i,v1} 
	 \fmf{vanilla,tension=0.12}{v2,bf,v1} 
	 \fmf{vanilla,tension=3.}{v2,o}
	 \fmf{photon,left,tension=1.}{v1,v2}
	 \fmfblob{.12w}{bf}
	 \fmfblob{.12w}{bb}
	 \fmfblob{.12w}{v2}
	 \fmfdot{v1}	 
	\end{fmfgraph*}
} \nonumber \hspace*{5mm}, \\
 \hspace*{15mm}{\cal S}^{-1}  &=& \hspace*{15mm} {\cal S}_0^{-1} \hspace*{18mm}+ \hspace*{12mm}\Sigma
\label{DSE:electron}
\end{eqnarray}
for fermions and
\begin{eqnarray}
\left[
\parbox{30mm}
{
	\begin{fmfgraph*}(80,40)
	 \fmfleft{i}\fmfright{o}
	 \fmf{photon,tension=3.}{i,b,o}
	 \fmfblob{.12w}{b}
	\end{fmfgraph*}
}
\right]^{-1} &=&
\left[
\parbox{30mm}
{
	\begin{fmfgraph*}(80,40)
	 \fmfleft{i}\fmfright{o}
	 \fmf{photon,tension=3.}{i,b,o}
	\end{fmfgraph*}
}
\right]^{-1} +
\parbox{25mm}
{
	\begin{fmfgraph*}(80,37)
	 \fmfleft{i}\fmfright{o}\fmftop{b1}\fmfbottom{b2}
	 \fmf{photon,tension=1.}{i,v1} 
	 \fmf{vanilla,tension=0.28,left}{v2,v1,v2} 
	 \fmf{photon,tension=1.}{v2,o}
	 \fmfdot{v1}
	 \fmfblob{.12w}{v2}
	 \fmfblob{.12w}{b1,b2}	 
	\end{fmfgraph*}
} \nonumber  \hspace*{5mm}, \\
\hspace*{12mm}({\cal D}^{-1})_{\mu\nu}  &=& \hspace*{13mm} ({\cal D}_0^{-1})_{\mu\nu} \hspace*{13mm} + \hspace*{13mm} \Pi_{\mu\nu} \label{DSE:photon}
\end{eqnarray}
\end{fmffile}
for bosons,
where the self-energies $\Sigma$ and $\Pi_{\mu\nu}$ are functionals of the dressed propagators $\cal S$ and ${\cal D}_{\mu\nu}$ (indicated by fat blobs) as well as of the dressed vertex-function $\Gamma_\mu$ (also fat blobs). Thus, these expressions couple to higher order DSE for higher n-point functions.

In order to calculate the asymptotic masses from the DSE, we use momentum independent asymptotic thermal mass expressions as ansatz for dressing the propagators, because the asymptotic thermal masses result from mutual scatterings among hard plasma particles with no coupling to the soft momentum excitations.
(This ansatz is in an analog spirit as the approach in~\cite{KarschPetreczkyPatkos}.)
Then, the propagators read in Feynman gauge
\begin{eqnarray}
 {\cal S}&=& \frac{1}{\feyndagg{K} - m_F}, \\
 {\cal D}_{\mu\nu} &=& \frac{{\mathfrak P}_{\mu\nu}^T}{K^2 - m_B^2}+\frac{g_{\mu\nu}-{\mathfrak P}_{\mu\nu}^T }{K^2},
\end{eqnarray}
where we only dressed the transverse part of the gauge boson propagator~\cite{RebhanFlechsig}.
The Ward-identity, $\partial {\cal S}^{-1} / \partial K_\mu = \gamma^\mu$ suggests to use the bare vertex, giving rise to a natural decoupling from the higher order DSE.
We focus, here, on the poles of these propagators at hard momenta by solving the DSE in (\ref{DSE:electron}) and (\ref{DSE:photon}) in the asymptotic region.
As a result, by using only the temperature dependent parts of the self-energies,
we find for the gap-equations

\begin{figure}[t]
\include{small_frag}
\psfrag{xlabelB}{\hspace*{1mm}\small $m/T$}
\psfrag{ylabelB}{\hspace*{-8mm} \small $m_B^2 / m_B^2({\rm HTL})$}
\psfrag{xlabelF}{\hspace*{1mm}\small $m/T$}
\psfrag{ylabelF}{\hspace*{-15mm} \small $(m_F^2 - m^2) / m_F^2({\rm HTL})$}
\psfrag{line1}{\tiny $g = 0.03$}
\psfrag{line2}{\tiny $g = 0.3$}
\psfrag{line3}{\tiny $g = 1.0$}
\psfrag{line4}{\tiny $g = 3.0$}
\psfrag{line1loop}{\tiny \rm 1loop}
\includegraphics[scale=0.32,angle=-90]{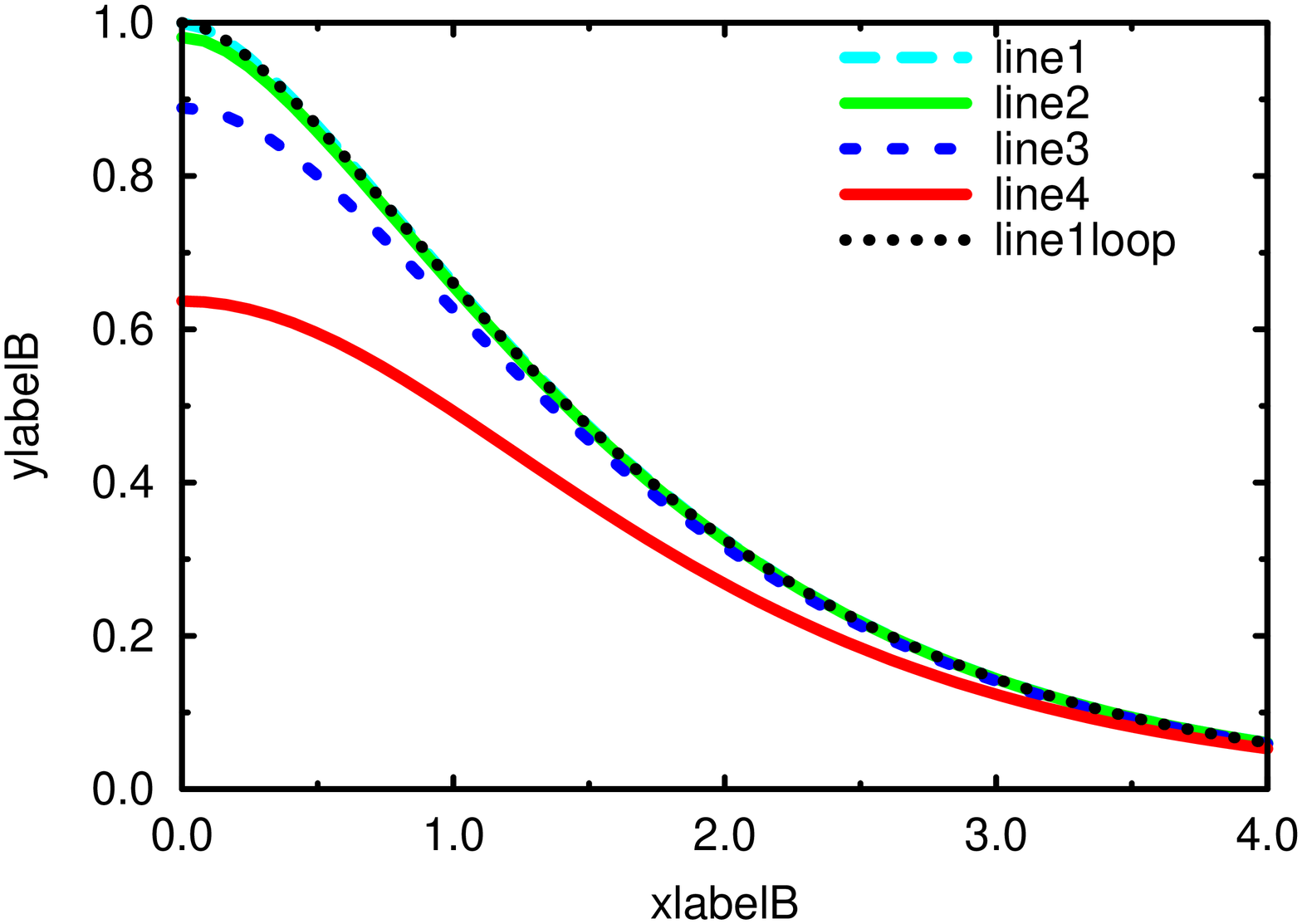}
\includegraphics[scale=0.32,angle=-90]{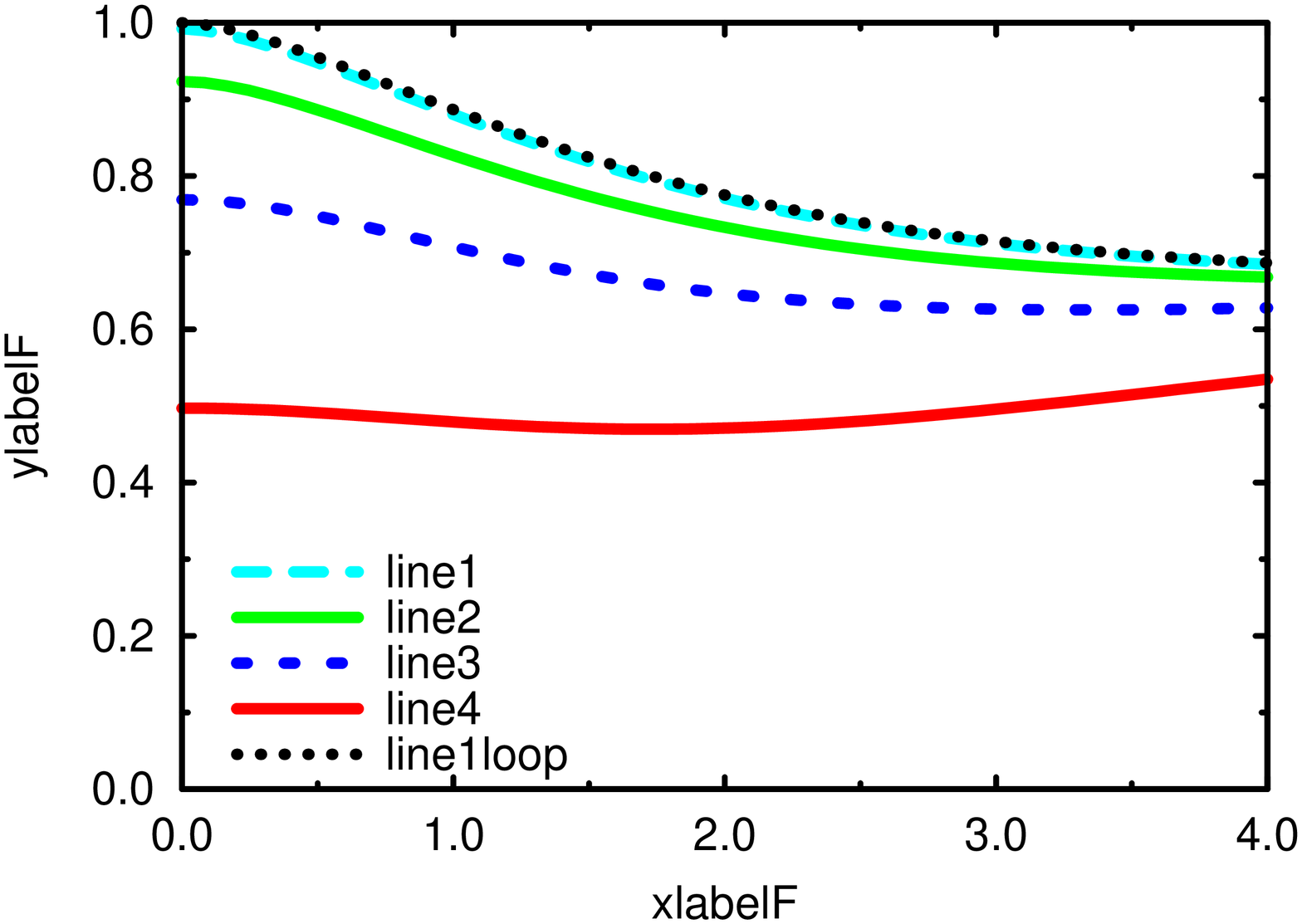}
\caption{The mass dependence of the scaled asymptotic thermal masses $m_B$ and $m_F$ for Abelian gauge bosons (left panel) and fermions (right panel), respectively. The self-consistent solutions of the gap-equations (\ref{DSE:F}) and (\ref{DSE:B}) are represented by solid and dashed lines for various values of the coupling $g$. The black dotted lines show the one-loop results given in (\ref{eq:asy_gluon_mass}) and (\ref{M_plus}), respectively,
using the substitutions described in the text to go from non-Abelian (QCD) to Abelian (QED) gauge theory. In the notation of the preceding sections this would be $m_\infty^2/m_g^2$ for the gauge boson and $M_+^2/(2m_f^2)$ for the fermion.
  \label{fig.9}}
\end{figure}

\begin{eqnarray}
 m_F^2 &=& m^2 + \frac{g^2}{12}(2{\cal I}_B(m_B) + {\cal I}_F(m_F)), \label{DSE:F}\\
 m_B^2 &=& \frac{g^2}{6} {\cal I}_F(m_F) \label{DSE:B}
\end{eqnarray}
with
\begin{eqnarray}
 {\cal I}_F &=& \frac{12}{\pi^2} \int \limits_0^\infty d k \frac{k^2}{\epsilon_F} n_F(\epsilon_F), \\
 {\cal I}_B &=& \frac{6}{\pi^2}  \int \limits_0^\infty d k \frac{k^2}{\epsilon_B} n_B(\epsilon_B),
\end{eqnarray}
where $\epsilon_B^2 = k^2 + m_B^2$, $\epsilon_F^2 = k^2 + m_F^2$. Self-consistent solutions of these equations
for $m_F^2$ and $m_B^2$ are presented in figure \ref{fig.9}.
For small values of the coupling, say $g=0.03$ (cyan long-dashed curves), the self-consistent solutions confirm the mass dependence in the one-loop results quite well. For larger values of the coupling, some deviations between the two approaches emanate, in particular in the fermionic sector, where already for $g \geq 0.3$ the deviation becomes
visible. In the strong coupling regime ($g > 1$) the difference becomes 50\% and larger. The coupling dependence
of the self-consistent fermionic mass at $m=0$ shows a similar behaviour as was found in~\cite{Nakkagawa} insofar, as in both approaches the self-consistent mass is much smaller than the corresponding HTL or one-loop values,
although~\cite{Nakkagawa} calculated the long-wavelength limit of the fermion dispersion relation while we concentrated on the asymptotic limit.

\end{document}